\documentclass[aps,prb,twocolumn,superscriptaddress]{revtex4}
\usepackage{amssymb,amsbsy,graphicx,times,subfigure,marvosym,color,bm,multirow}
\usepackage[latin1]{inputenc}
\begin{document}

\title{Coherent hole propagation in an exactly solvable gapless spin liquid}

\author{G\'abor B. Hal\'asz}
\affiliation{Kavli Institute for Theoretical Physics, University of
California, Santa Barbara, CA 93106, USA}
\affiliation{Theoretical
Physics, Oxford University, 1 Keble Road, Oxford OX1 3NP, United
Kingdom}

\author{J. T. Chalker}
\affiliation{Theoretical Physics, Oxford University, 1 Keble Road,
Oxford OX1 3NP, United Kingdom}


\begin{abstract}

We examine the dynamics of a single hole in the gapless phase of the
Kitaev honeycomb model, focusing on the slow-hole regime where the
bare hopping amplitude $t$ is much less than the Kitaev exchange
energy $J$. In this regime, the hole does not generate gapped flux
excitations and is dressed only by the gapless fermion excitations.
Investigating the single-hole spectral function, we find that the
hole propagates coherently with a quasiparticle weight that is
finite but approaches zero as $t/J \to 0$. This conclusion follows
from two approximate treatments, which capture the same physics in
complementary ways. Both treatments use the stationary limit as an
exactly solvable starting point to study the spectral function
approximately (i) by employing a variational approach in terms of a
trial state that interpolates between the limits of a stationary
hole and an infinitely fast hole and (ii) by considering a special
point in the gapless phase that corresponds to a simplified
one-dimensional problem.

\end{abstract}


\maketitle


\section{Introduction} \label{sec-int}

The physics of a doped Mott insulator is a central problem in the
field of strongly correlated electrons,\cite{Mott} being motivated
in part by high-temperature superconductivity in the
cuprates.\cite{Anderson-0} One main question is whether the charge
carriers (electrons or holes) form a Fermi liquid or a non-Fermi
liquid in a lightly doped Mott insulator.\cite{Anderson-1} In a
Fermi liquid, charge carriers propagate as coherent quasiparticles,
which is indicated by a corrresponding delta-function peak in the
single-particle spectral function. Conversely, non-Fermi liquids are
characterized by a completely incoherent propagation of charge
carriers.\cite{Schofield} Due to their different spectral functions,
one can distinguish these two possibilities by using angle-resolved
photoemission spectroscopy.\cite{Damascelli} Moreover, Fermi liquids
and non-Fermi liquids exhibit different thermodynamic and transport
properties at low temperatures.\cite{Stewart}

Since the ground state of a stereotypical Mott insulator is
antiferromagnetically (AFM) ordered, the standard description of a
lightly doped Mott insulator is in terms of an appropriate $t$-$J$
model with AFM Heisenberg interactions.\cite{t-J} It has been
established that, in two dimensions, a single hole propagates
coherently in such an AFM ordered state.\cite{AFM-1} However, it is
also known that the ground state of a Mott insulator does not have
to be AFM ordered or even magnetically ordered at
all.\cite{Misguich} In particular, Anderson
suggested\cite{Anderson-2} that the parent state of a
high-temperature superconductor is a quantum spin
liquid,\cite{Balents} an exotic strongly-correlated state exhibiting
long-range entanglement,\cite{Savary} fractional
excitations,\cite{Rajaraman} and a topological ground-state
degeneracy.\cite{Wen} The melting of the AFM order into such a
spin-liquid state is particularly favored by doping as the holes can
then propagate more freely without scrambling an underlying magnetic
order.\cite{Brinkman, AFM-2} Nevertheless, it is far from obvious
whether a single hole in a spin liquid propagates as a coherent
quasiparticle.

\begin{figure}[h]
\centering
\includegraphics[width=0.47\columnwidth]{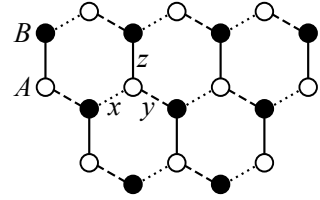}
\caption{Illustration of the honeycomb lattice. Sites in sublattices
$A$ and $B$ are marked by white and black circles, while $x$, $y$,
and $z$ bonds are marked by dotted, dashed, and solid lines,
respectively.} \label{fig-1}
\end{figure}

In this work, we address this challenging question for the Kitaev
honeycomb model, an exactly solvable yet realistic spin model with a
spin-liquid ground state.\cite{Kitaev} This model consists of $S =
1/2$ spins at the sites of a honeycomb lattice, which are coupled
via different spin components along the three bonds connected to any
given site. The Hamiltonian is
\begin{equation}
H_K = -J_x \sum_{\langle \mathbf{r}, \mathbf{r}' \rangle_x}
\sigma_{\mathbf{r}}^x \sigma_{\mathbf{r}'}^x - J_y \sum_{\langle
\mathbf{r}, \mathbf{r}' \rangle_y} \sigma_{\mathbf{r}}^y
\sigma_{\mathbf{r}'}^y - J_z \sum_{\langle \mathbf{r}, \mathbf{r}'
\rangle_z} \sigma_{\mathbf{r}}^z \sigma_{\mathbf{r}'}^z,
\label{eq-int-H}
\end{equation}
where $J_{x,y,z}$ are the coupling constants for the three types of
bonds $x$, $y$, and $z$ (see Fig.~\ref{fig-1}). Depending on these
coupling constants, the ground state is either a gapped or a gapless
spin liquid. In an earlier work,\cite{Halasz-1} we provided a
systematic study of slow-hole dynamics in the gapped phase of the
model, discussing the single-particle properties (e.g., particle
statistics) of individual holes and describing two different
(fractional) Fermi-liquid ground states at finite doping. Due to the
absence of low-energy excitations, slow holes in the gapped phase
are necessarily coherent quasiparticles. In the present work, we
focus on the gapless phase and investigate whether a single hole
propagates coherently. The answer to this question is one step
towards understanding whether the holes form a Fermi liquid at a
small but finite density.

This work complements several papers in the existing literature.
First, the lightly doped Kitaev honeycomb model has been studied
extensively in the framework of slave-particle mean-field
theories.\cite{MFT, Mei, You} Although this approach provides a full
classification of spin-liquid ground states consistent with a given
set of symmetries,\cite{PSG} it is not immediately clear which one
of these ground states is actually realized for a particular
Hamiltonian. Indeed, the conclusions of Refs.~\onlinecite{Mei} and
\onlinecite{You} disagree as the former predicts a Fermi-liquid
state and the latter predicts a non-Fermi-liquid state at small
doping. Second, a single hole in the Kitaev honeycomb model has been
studied in Ref.~\onlinecite{Trousselet} via exact diagonalization of
small systems. Within the limits of their calculation, the authors
find that a fast hole with hopping amplitude $t \gtrsim J_{x,y,z}$
propagates incoherently. Our study is complementary to theirs in two
ways as we consider a slow hole with hopping amplitude $t \ll
J_{x,y,z}$ and employ the exact solution of the model to obtain
analytic results that are applicable in the thermodynamic limit.

The main result of this paper is that a slow hole in the gapless
phase of the Kitaev honeycomb model propagates as a coherent
quasiparticle. Indeed, the single-hole spectral function is found to
have a low-energy delta-function peak. The quasiparticle weight, the
coefficient of this delta-function peak, is finite for any hopping
amplitude $t > 0$ but vanishes in the stationary limit $t
\rightarrow 0$. Since the model is no longer exactly solvable in the
presence of a mobile hole, we deal with the problem approximately by
using two complementary directions. First, we employ a variational
approach in terms of a single-parameter trial state that
interpolates smoothly between the extreme limits of a stationary
hole and an infinitely fast hole. Second, we consider a simplified
one-dimensional problem that captures the low-energy physics at a
special point in the gapless phase. The results from these two
directions are fully consistent with each other and strongly
corroborate our claims on coherent propagation.

The paper is structured as follows. In Sec.~\ref{sec-gen}, we
introduce the problem in a convenient formalism that is used
throughout the rest of the paper. In Sec.~\ref{sec-stat}, we
consider the exactly solvable limit of a stationary hole as a
starting point of our investigation. In Secs.~\ref{sec-var} and
\ref{sec-1D}, we discuss the two complementary directions for
treating a mobile hole, the variational approach and the simplified
one-dimensional problem, respectively. In Sec.~\ref{sec-disc}, we
compare the results from these two directions and also those from
previous works. Finally, in Sec.~\ref{sec-out}, we conclude the
paper with suggestions for future research.

\section{General formulation} \label{sec-gen}

In the most general case, the lightly doped Kitaev honeycomb model
is described by a modified $t$-$J$ model\cite{t-J} where the usual
Heisenberg interactions are substituted with the Kitaev couplings in
Eq.~(\ref{eq-int-H}). The Hamiltonian of this model reads
\begin{equation}
H = H_K - t \sum_{\langle \mathbf{r}, \mathbf{r}' \rangle}
\sum_{\sigma} \left( \mathcal{P} a_{\mathbf{r}, \sigma}^{\dag}
a_{\mathbf{r}', \sigma}^{\phantom{\dag}} \mathcal{P} + \mathrm{H.c.}
\right), \label{eq-gen-H-1}
\end{equation}
where $a_{\mathbf{r}, \sigma}^{\dag}$ creates an electron with spin
$\sigma$ at site $\mathbf{r}$, and $\mathcal{P}$ projects out states
with double occupancy. Formally, the spin operators in $H_K$ are
expressed as $\sigma_{\mathbf{r}}^{\alpha} = a_{\mathbf{r},
\sigma_1}^{\dag} \tau_{\sigma_1, \sigma_2}^{\alpha} a_{\mathbf{r},
\sigma_2}^{\phantom{\dag}}$ in terms of the electron operators,
where $\tau^{\alpha}$ are the Pauli matrices with $\alpha = \{ x,y,z
\}$. Our main quantity of interest, the single-hole spectral
function, is then given by
\begin{equation}
\mathcal{A} (\varepsilon, \mathbf{K}) = \sum_{\lambda} \sum_{\sigma}
\left| \big{\langle} \tilde{\Phi}_{\lambda} \big|
\hat{a}_{-\mathbf{K}, \sigma} \big| \Omega \big{\rangle} \right|^2
\delta \big[ \varepsilon - \tilde{E}_{\lambda} \big],
\label{eq-gen-A-1}
\end{equation}
where $| \Omega \rangle$ is the ground state of the model without
any holes (undoped model), $| \tilde{\Phi}_{\lambda} \rangle$ are
the eigenstates of the model with a single hole (doped model), and
$\hat{a}_{-\mathbf{K}, \sigma} \propto \sum_{\mathbf{r}} e^{i
\mathbf{K} \cdot \mathbf{r}} a_{\mathbf{r}, \sigma}$ creates a hole
with momentum $\mathbf{K}$. Since $\tilde{E}_{\lambda}$ is the
energy of the eigenstate $| \tilde{\Phi}_{\lambda} \rangle$, the
spectral function $\mathcal{A} (\varepsilon, \mathbf{K})$ is the
energy distribution of the single-hole state $\hat{a}_{-\mathbf{K},
\sigma} | \Omega \rangle$. Note that we consistently use a tilde to
distinguish quantities of the doped model from those of the undoped
model.

Restricting our attention to a single hole in the model, we do not
consider the general Hamiltonian in Eq.~(\ref{eq-gen-H-1}) but
describe the mobile hole in first quantization instead. The
Hamiltonian of the undoped model is simply $H_{\sigma} \equiv H_K$
in terms of the spin degrees of freedom, while the doped model
contains an additional degree of freedom specifying the hole
position in the lattice (i.e., the hole site). Furthermore, we
account for the presence of the hole via the hole-spin picture used
in Ref.~\onlinecite{Halasz-1}. Instead of actually removing the spin
from the hole site, we switch off its couplings to all the other
spins. To avoid introducing an unphysical degeneracy, we may demand
that this hole spin is always in the spin-up state. In terms of the
hole hopping amplitude $t$, the block of the doped Hamiltonian
connecting hole sites $\mathbf{r}_1$ and $\mathbf{r}_2$ is then
\begin{eqnarray}
\tilde{H}_{\sigma} (\mathbf{r}_1, \mathbf{r}_2) &=& \left[
H_{\sigma} + \sum_{\alpha} J_{\alpha}^{\phantom{\alpha}}
\sigma_{\mathbf{r}_1}^{\alpha} \sigma_{\mathbf{r}_1 \pm
\hat{\mathbf{r}}_{\alpha}}^{\alpha} \right] \delta_{\mathbf{r}_1,
\mathbf{r}_2}
\nonumber \\
&& -\frac{t}{2} \left[ 1 + \bm{\sigma}_{\mathbf{r}_1} \cdot
\bm{\sigma}_{\mathbf{r}_2} \right] \, \sum_{\alpha}
\delta_{\mathbf{r}_1 \pm \hat{\mathbf{r}}_{\alpha}, \mathbf{r}_2},
\qquad \label{eq-gen-H-2}
\end{eqnarray}
where $\hat{\mathbf{r}}_{\alpha}$ is the vector along an $\alpha$
bond from a site in sublattice $A$ to a neighboring site in
sublattice $B$ (see Fig.~\ref{fig-1}), and the upper (lower) sign in
front of $\hat{\mathbf{r}}_{\alpha}$ corresponds to $\mathbf{r}_1
\in A$ ($\mathbf{r}_1 \in B$). The terms $J_{\alpha}
\sigma_{\mathbf{r}_1}^{\alpha} \sigma_{\mathbf{r}_1 \pm
\hat{\mathbf{r}}_{\alpha}}^{\alpha}$ in the diagonal blocks describe
the switched-off couplings around the hole site, while the operators
$[1 + \bm{\sigma}_{\mathbf{r}_1} \cdot \bm{\sigma}_{\mathbf{r}_2}] /
2$ in the off-diagonal blocks exchange the hole spin with one of its
neighbors.

In its gapless phase, the elementary excitations of the Kitaev model
are gapless fermions and gapped fluxes. Since we are interested in
the low-energy physics for a small hopping amplitude $t \ll
J_{\alpha}$ and vanishing hole density, we neglect the flux
excitations and consider the interplay between the mobile hole and
the fermion excitations only. Employing the exact solution of the
model in the standard way,\cite{Kitaev} and restricting our
attention to the low-energy sector with no flux excitations, we end
up with one Majorana fermion $\hat{c}_{\mathbf{r}}$ at each site
$\mathbf{r}$, and an effective Hamiltonian in terms of these
Majorana fermions. The details of this procedure are explained in
Appendix \ref{app-1}. For the undoped model, the effective
low-energy Hamiltonian is
\begin{equation}
H_c = \sum_{\alpha} \sum_{\mathbf{r} \in A} i J_{\alpha}
\hat{c}_{\mathbf{r}} \hat{c}_{\mathbf{r} +
\hat{\mathbf{r}}_{\alpha}}, \label{eq-gen-H-3}
\end{equation}
while for the doped model, its respective blocks are
\begin{eqnarray}
\tilde{H}_c (\mathbf{r}_1, \mathbf{r}_2) &=& \left[ H_c \mp
\sum_{\alpha} i J_{\alpha} \hat{c}_{\mathbf{r}_1}
\hat{c}_{\mathbf{r}_1 \pm \hat{\mathbf{r}}_{\alpha}} \right]
\delta_{\mathbf{r}_1, \mathbf{r}_2}
\nonumber \\
&& -\frac{t}{2} \left[ 1 \mp i \hat{c}_{\mathbf{r}_1}
\hat{c}_{\mathbf{r}_2} \right] \, \sum_{\alpha} \delta_{\mathbf{r}_1
\pm \hat{\mathbf{r}}_{\alpha}, \mathbf{r}_2}, \label{eq-gen-H-4}
\end{eqnarray}
where the upper (lower) sign again corresponds to $\mathbf{r}_1 \in
A$ ($\mathbf{r}_1 \in B$). The undoped Hamiltonian consists of
quadratic coupling terms between neighboring Majorana fermions. In
the diagonal blocks of the doped Hamiltonian, these coupling terms
are switched off around the hole site.

Since the low-energy fermions are perturbed by the presence of the
hole in the doped model, it is useful to relabel the Majorana
fermions $\hat{c}_{\mathbf{r}}$ by their relative positions with
respect to the hole site $\mathbf{r}_0$. Taking a reference site
$\mathbf{0} \in A$, the Majorana fermions are consistently relabeled
as
\begin{eqnarray}
&& \hat{c}_{\mathbf{r}} \rightarrow c_{\mathbf{r} - \mathbf{r}_0}
\qquad (\mathbf{r}_0 \in A),
\nonumber \\
&& \hat{c}_{\mathbf{r}} \rightarrow c_{\mathbf{r}_0 - \mathbf{r}}
\qquad (\mathbf{r}_0 \in B, \mathbf{r} \in A),
\label{eq-gen-c} \\
&& \hat{c}_{\mathbf{r}} \rightarrow -c_{\mathbf{r}_0 - \mathbf{r}}
\quad \, (\mathbf{r}_0 \in B, \mathbf{r} \in B). \nonumber
\end{eqnarray}
In the case of $\mathbf{r}_0 \in A$, this relabeling corresponds to
a translation, while in the case of $\mathbf{r}_0 \in B$, it
corresponds to an inversion exchanging the two sublattices.

In terms of the relabeled Majorana fermions $c_{\mathbf{r}}$, the
undoped Hamiltonian in Eq.~(\ref{eq-gen-H-3}) is then
\begin{equation}
H_c = \sum_{\alpha} \sum_{\mathbf{r} \in A} i J_{\alpha}
c_{\mathbf{r}} c_{\mathbf{r} + \hat{\mathbf{r}}_{\alpha}},
\label{eq-gen-H-5}
\end{equation}
while the blocks of the doped Hamiltonian in Eq.~(\ref{eq-gen-H-4})
are
\begin{eqnarray}
\tilde{H}_c (\mathbf{r}_1, \mathbf{r}_2) &=& \left[ H_c -
\sum_{\alpha} i J_{\alpha} c_{\mathbf{0}}
c_{\hat{\mathbf{r}}_{\alpha}} \right] \delta_{\mathbf{r}_1,
\mathbf{r}_2}
\label{eq-gen-H-6} \\
&& -\frac{t}{2} \sum_{\alpha} \left[ \hat{R}_{\alpha} - i
c_{\mathbf{0}} \hat{R}_{\alpha} c_{\mathbf{0}} \right]
\delta_{\mathbf{r}_1 \pm \hat{\mathbf{r}}_{\alpha}, \mathbf{r}_2}.
\nonumber
\end{eqnarray}
In each off-diagonal block of the doped Hamiltonian, the two
relabeling conventions for the two neighboring hole sites must be
related by an appropriate operator $\hat{R}_{\alpha}$ that
corresponds to an inversion $R_{\alpha}$ around the center of the
$\alpha$ bond connecting the two sites. We express this inversion
operator $\hat{R}_{\alpha}$ via the fermions that diagonalize the
undoped Hamiltonian in Eq.~(\ref{eq-gen-H-5}). Since these fermions
are labeled by their momenta $\mathbf{k}$ due to translation
symmetry, and those with momenta $\pm \mathbf{k}$ are degenerate due
to inversion symmetry, we can define appropriate even ($\eta$) and
odd ($\mu$) complex fermions
\begin{eqnarray}
\psi_{\mathbf{k}, \eta} (\alpha) &=& \frac{1}{2} \left[
\gamma_{\mathbf{k}, \eta, A} (\alpha) + i \gamma_{\mathbf{k}, \eta,
B} (\alpha) \right],
\nonumber \\
\psi_{\mathbf{k}, \mu} (\alpha) &=& \frac{1}{2} \left[
\gamma_{\mathbf{k}, \mu, A} (\alpha) + i \gamma_{\mathbf{k}, \mu, B}
(\alpha) \right] \label{eq-gen-psi-1}
\end{eqnarray}
such that their Majorana fermion components
\begin{eqnarray}
\gamma_{\mathbf{k}, \eta, \Xi} (\alpha) &\propto& \sum_{\mathbf{r}
\in \Xi} \cos \left[ \mathbf{k} \cdot \left( \mathbf{r} -
\hat{\mathbf{r}}_{\alpha} / 2 \right) \right] c_{\mathbf{r}},
\nonumber \\
\gamma_{\mathbf{k}, \mu, \Xi} (\alpha) &\propto& \sum_{\mathbf{r}
\in \Xi} \sin \left[ \mathbf{k} \cdot \left( \mathbf{r} -
\hat{\mathbf{r}}_{\alpha} / 2 \right) \right] c_{\mathbf{r}}
\label{eq-gen-gamma-1}
\end{eqnarray}
corresponding to the two sublattices $\Xi = A,B$ have even ($\eta$)
and odd ($\mu$) envelope functions with respect to the center of the
$\alpha$ bond. Under the inversion $R_{\alpha}$, these Majorana
fermion components transform as
\begin{eqnarray}
\hat{R}_{\alpha} \left[ \gamma_{\mathbf{k}, \eta, A} (\alpha)
\right] &=& \gamma_{\mathbf{k}, \eta, B} (\alpha),
\nonumber \\
\hat{R}_{\alpha} \left[ \gamma_{\mathbf{k}, \eta, B} (\alpha)
\right] &=& -\gamma_{\mathbf{k}, \eta, A} (\alpha),
\nonumber \\
\hat{R}_{\alpha} \left[ \gamma_{\mathbf{k}, \mu, A} (\alpha) \right]
&=& -\gamma_{\mathbf{k}, \mu, B} (\alpha),
\label{eq-gen-gamma-2} \\
\hat{R}_{\alpha} \left[ \gamma_{\mathbf{k}, \mu, B} (\alpha) \right]
&=& \gamma_{\mathbf{k}, \mu, A} (\alpha), \nonumber
\end{eqnarray}
and therefore the complex fermions transform as
\begin{eqnarray}
\hat{R}_{\alpha} \left[ \psi_{\mathbf{k}, \eta} (\alpha) \right] &=&
-i \psi_{\mathbf{k}, \eta} (\alpha),
\nonumber \\
\hat{R}_{\alpha} \left[ \psi_{\mathbf{k}, \mu} (\alpha) \right] &=&
i \psi_{\mathbf{k}, \mu} (\alpha). \label{eq-gen-psi-2}
\end{eqnarray}
Since the vacuum state $| \omega \rangle$ of these fermions [i.e.,
the ground state of the undoped Hamiltonian in
Eq.~(\ref{eq-gen-H-5})] is invariant under $R_{\alpha}$, the
inversion operator is then given by
\begin{eqnarray}
\hat{R}_{\alpha} &=& \exp \left\{ \frac{i \pi} {2} \sum_{\pm
\mathbf{k}} \left[ \psi_{\mathbf{k}, \eta}^{\dag} (\alpha) \,
\psi_{\mathbf{k}, \eta}^{\phantom{\dag}} (\alpha) -
\psi_{\mathbf{k}, \mu}^{\dag} (\alpha) \, \psi_{\mathbf{k},
\mu}^{\phantom{\dag}} (\alpha) \right] \right\}
\nonumber \\
&=& \prod_{\pm \mathbf{k}} \bigg\{ \left[ 1 - (1 - i) \,
\psi_{\mathbf{k}, \eta}^{\dag} (\alpha) \, \psi_{\mathbf{k},
\eta}^{\phantom{\dag}} (\alpha) \right]
\label{eq-gen-R} \\
&& \times \left[ 1 - (1 + i) \, \psi_{\mathbf{k}, \mu}^{\dag}
(\alpha) \, \psi_{\mathbf{k}, \mu}^{\phantom{\dag}} (\alpha) \right]
\bigg\} \, , \nonumber
\end{eqnarray}
where $\pm \mathbf{k}$ corresponds to pairs of momenta. Since the
unitary operator $\hat{R}_{\alpha}$ is Hermitian for even fermion
number and anti-Hermitian for odd fermion number, it is effectively
Hermitian because the fermion number is always even for physical
states in the zero-flux sector of the Kitaev model.\cite{Kitaev}

While the eigenstates of the undoped Hamiltonian in
Eq.~(\ref{eq-gen-H-5}) belong to fermion space only, those of the
doped Hamiltonian in Eq.~(\ref{eq-gen-H-6}) belong to the product of
fermion space and hole position space. If we assume that they do not
break translation or inversion symmetry, these eigenstates can be
written as
\begin{equation}
\big| \tilde{\theta}_{\mathbf{K}} \big{\rangle} \propto \left[
\sum_{\mathbf{r} \in A} e^{i \mathbf{K} \cdot \mathbf{r}} |
\mathbf{r} \rangle + \sum_{\mathbf{r} \in B} e^{i \mathbf{K} \cdot
\mathbf{r} + i \tilde{\vartheta}_{\mathbf{K}}} | \mathbf{r} \rangle
\right] \otimes \big| \tilde{\chi}_{\mathbf{K}}^{\phantom{\dag}}
\big{\rangle}, \label{eq-gen-theta}
\end{equation}
where $| \tilde{\chi}_{\mathbf{K}}^{\phantom{\dag}} \rangle$ is a
state in fermion space, and $| \mathbf{r} \rangle$ is a state in
hole position space corresponding to hole site $\mathbf{r}$.
Translation symmetry gives rise to a hole momentum $\mathbf{K}$,
while inversion symmetry gives rise to a phase difference
$\tilde{\vartheta}_{\mathbf{K}}$ between the two sublattices.
Substituting Eq.~(\ref{eq-gen-theta}) into Eq.~(\ref{eq-gen-H-6}),
we obtain that $| \tilde{\chi}_{\mathbf{K}}^{\phantom{\dag}}
\rangle$ are eigenstates of the effective Hamiltonian
\begin{eqnarray}
\tilde{H}_c (\mathbf{K}) &=& H_c - \sum_{\alpha} i J_{\alpha}
c_{\mathbf{0}} c_{\hat{\mathbf{r}}_{\alpha}}
\label{eq-gen-H-7} \\
&& \, -\frac{t}{2} \sum_{\alpha} \left[ \hat{R}_{\alpha} - i
c_{\mathbf{0}} \hat{R}_{\alpha} c_{\mathbf{0}} \right] \cos \big[
\mathbf{K} \cdot \hat{\mathbf{r}}_{\alpha} +
\tilde{\vartheta}_{\mathbf{K}} \big], \nonumber
\end{eqnarray}
where the phase difference $\tilde{\vartheta}_{\mathbf{K}}$ is in
general determined self-consistently for each eigenstate by
\begin{equation}
\sum_{\alpha} \sin \big[ \mathbf{K} \cdot \hat{\mathbf{r}}_{\alpha}
+ \tilde{\vartheta}_{\mathbf{K}} \big] \, \big{\langle}
\tilde{\chi}_{\mathbf{K}}^{\phantom{\dag}} \big| \left[
\hat{R}_{\alpha} - i c_{\mathbf{0}} \hat{R}_{\alpha} c_{\mathbf{0}}
\right] \big| \tilde{\chi}_{\mathbf{K}}^{\phantom{\dag}}
\big{\rangle} = 0. \label{eq-gen-vartheta}
\end{equation}
In the special case of zero hole momentum $\mathbf{K} = \mathbf{0}$,
it is either $\tilde{\vartheta}_{\mathbf{0}} = 0$ or
$\tilde{\vartheta}_{\mathbf{0}} = \pi$ for all eigenstates $|
\tilde{\chi}_{\mathbf{0}}^{\phantom{\dag}} \rangle$. The main
advantage of Eq.~(\ref{eq-gen-H-7}) with respect to
Eq.~(\ref{eq-gen-H-6}) is that its eigenstates $|
\tilde{\chi}_{\mathbf{K}}^{\phantom{\dag}} \rangle$ belong to
fermion space only and are therefore directly comparable to the
eigenstates $| \chi \rangle$ of the undoped Hamiltonian in
Eq.~(\ref{eq-gen-H-5}). In particular, the single-hole spectral
function can be expressed in terms of these eigenstates as
\begin{eqnarray}
\mathcal{A} (\varepsilon, \mathbf{K}) &=& \frac{1}{2}
\sum_{\lambda_{\mathbf{K}}} \left( 1 + \cos
\tilde{\vartheta}_{\mathbf{K}, \lambda_{\mathbf{K}}} \right) \left|
\big{\langle} \tilde{\chi}_{\mathbf{K},
\lambda_{\mathbf{K}}}^{\phantom{\dag}} \big| \omega \big{\rangle}
\right|^2 \nonumber \\
&& \times \, \delta \left[ \varepsilon - \tilde{E}_{\mathbf{K},
\lambda_{\mathbf{K}}} \right], \label{eq-gen-A-2}
\end{eqnarray}
where $\tilde{E}_{\mathbf{K}, \lambda_{\mathbf{K}}}$ is the energy
of the eigenstate $| \tilde{\chi}_{\mathbf{K},
\lambda_{\mathbf{K}}}^{\phantom{\dag}} \rangle$, and
$\lambda_{\mathbf{K}}$ is an additional label to distinguish
eigenstates that correspond to the same hole momentum $\mathbf{K}$.
We provide a detailed derivation of this result in Appendix
\ref{app-2}.

\section{Stationary limit} \label{sec-stat}

As a starting point of our investigation, we first consider the
stationary limit ($t = 0$) when the undoped Hamiltonian $H_c$ in
Eq.~(\ref{eq-gen-H-5}) and the doped Hamiltonian $\tilde{H}_c \equiv
\tilde{H}_c (\mathbf{K})$ in Eq.~(\ref{eq-gen-H-7}) are both
quadratic and hence exactly solvable.\cite{Willans} For simplicity,
we also restrict our attention to the spatially isotropic point of
the model, at which $J_{x,y,z} = J_0$. Since we are interested in
the presence (or absence) of a delta-function peak in the spectral
function $\mathcal{A} (\varepsilon) \equiv \mathcal{A} (\varepsilon,
\mathbf{K})$, we aim to calculate the overlap $\langle
\tilde{\omega} | \omega \rangle$ between the undoped ground state $|
\omega \rangle$ and the doped ground state $| \tilde{\omega}
\rangle$. If this ground-state overlap is finite in the
thermodynamic limit, there is a delta-function peak in the spectral
function with a corresponding hole quasiparticle weight $Z = |
\langle \tilde{\omega} | \omega \rangle |^2 > 0$. Conversely, if the
ground-state overlap vanishes in the thermodynamic limit, the
presence of the hole leads to an orthogonality
catastrophe,\cite{Anderson-3} and the spectral function has no
delta-function peak ($Z = 0$).

To set up our calculation in a more standard formulation, we employ
a fermion doubling procedure, which turns our quadratic Majorana
fermion problems into quadratic number-conserving complex fermion
problems. We introduce Majorana fermion copies $c_{\mathbf{r}}'$ of
the original Majorana fermions $c_{\mathbf{r}}$, and define
corresponding complex fermions as
\begin{equation}
f_{\mathbf{r} \in A} = \frac{1}{2} \left( c_{\mathbf{r}} + i
c_{\mathbf{r}}' \right), \quad \,\, f_{\mathbf{r} \in B} =
\frac{i}{2} \left( c_{\mathbf{r}} + i c_{\mathbf{r}}' \right).
\label{eq-stat-f}
\end{equation}
For the undoped model, the doubled Hamiltonian is then
\begin{eqnarray}
H_c + H_c' &=& J_0 \sum_{\alpha} \sum_{\mathbf{r} \in A} \left( i
c_{\mathbf{r}} c_{\mathbf{r} + \hat{\mathbf{r}}_{\alpha}} + i
c_{\mathbf{r}}' c_{\mathbf{r} + \hat{\mathbf{r}}_{\alpha}}' \right)
\nonumber \\
&=& 2 J_0 \sum_{\alpha} \sum_{\mathbf{r} \in A} \left(
f_{\mathbf{r}}^{\dag} f_{\mathbf{r} +
\hat{\mathbf{r}}_{\alpha}}^{\phantom{\dag}} + f_{\mathbf{r} +
\hat{\mathbf{r}}_{\alpha}}^{\dag} f_{\mathbf{r}}^{\phantom{\dag}}
\right) \quad \label{eq-stat-H-1} \\
&\equiv& \sum_{\mathbf{r}, \mathbf{r}'} \mathcal{H}_{\mathbf{r},
\mathbf{r}'}^{\phantom{\dag}} f_{\mathbf{r}}^{\dag}
f_{\mathbf{r}'}^{\phantom{\dag}} \equiv f^{\dag} \cdot \mathcal{H}
\cdot f, \nonumber
\end{eqnarray}
while for the doped model, it takes the form
\begin{eqnarray}
\tilde{H}_c + \tilde{H}_c' &=& 2 J_0 \sum_{\alpha} \sum_{\mathbf{r}
\in A} \left( f_{\mathbf{r}}^{\dag} f_{\mathbf{r} +
\hat{\mathbf{r}}_{\alpha}}^{\phantom{\dag}} + f_{\mathbf{r} +
\hat{\mathbf{r}}_{\alpha}}^{\dag} f_{\mathbf{r}}^{\phantom{\dag}}
\right) \nonumber \\
&& \, - 2 J_0 \sum_{\alpha} \left( f_{\mathbf{0}}^{\dag}
f_{\hat{\mathbf{r}}_{\alpha}}^{\phantom{\dag}} +
f_{\hat{\mathbf{r}}_{\alpha}}^{\dag} f_{\mathbf{0}}^{\phantom{\dag}}
\right) \label{eq-stat-H-2} \\
&\equiv& \sum_{\mathbf{r}, \mathbf{r}'}
\tilde{\mathcal{H}}_{\mathbf{r}, \mathbf{r}'}^{\phantom{\dag}}
f_{\mathbf{r}}^{\dag} f_{\mathbf{r}'}^{\phantom{\dag}} \equiv
f^{\dag} \cdot \tilde{\mathcal{H}} \cdot f. \nonumber
\end{eqnarray}
The doubled Hamiltonian of the undoped model is identical to that of
graphene, and the additional term for the doped model corresponds to
an infinite potential introduced at site $\mathbf{0}$. Each doubled
Hamiltonian is particle-hole symmetric by construction and is
diagonalized by fermions that come in pairs with opposite (i.e.,
positive and negative) energies. In particular, the undoped model
has doubled fermions $\phi_{m, \pm}$ with energies $\epsilon_{m,
\pm} = \pm \epsilon_{m, +}$, and the doped model has doubled
fermions $\tilde{\phi}_{m, \pm}$ with energies $\tilde{\epsilon}_{m,
\pm} = \pm \tilde{\epsilon}_{m, +}$. The doubled ground state in
each case is then the state in which all of the negative-energy
fermions and none of the positive-energy fermions are excited.
Mathematically, these ground states read
\begin{eqnarray}
| \Omega' \rangle &=& | \omega \rangle \otimes | \omega' \rangle =
\prod_m \phi_{m,-}^{\dag} | 0 \rangle,
\nonumber \\
| \tilde{\Omega}' \rangle &=& | \tilde{\omega} \rangle \otimes |
\tilde{\omega}' \rangle = \prod_m \tilde{\phi}_{m,-}^{\dag} | 0
\rangle, \label{eq-stat-omega}
\end{eqnarray}
where $| 0 \rangle$ is the vacuum state of both the doubled fermions
$\phi_{m, \pm}$ and the doubled fermions $\tilde{\phi}_{m, \pm}$. In
the graphene language, the Fermi energy is at zero energy in both
cases, and all negative-energy levels are filled with particles.
However, the levels are perturbed by the infinite potential and, in
particular, there is mixing between the positive-energy and the
negative-energy levels. The perturbed ground state $|
\tilde{\Omega}' \rangle$ is therefore different from the unperturbed
one $| \Omega' \rangle$.

Using the doubled formulation, the quasiparticle weight is given by
$Z = | \langle \tilde{\omega} | \omega \rangle |^2 = \sqrt{ |
\langle \tilde{\Omega}' | \Omega' \rangle |^2 }$. If we define a
unitary matrix $\tilde{W}$ that transforms the perturbed fermions
$\tilde{\phi}_{m, \pm}$ into the unperturbed fermions $\phi_{m,
\pm}$ with the block structure
\begin{equation}
\left( \begin{array}{c} \phi_{+} \\ \phi_{-}
\end{array} \right) = \left( \begin{array}{cc} \tilde{W}_{+,+} &
\tilde{W}_{+,-} \\ \tilde{W}_{-,+} & \tilde{W}_{-,-} \end{array}
\right) \cdot \left(
\begin{array}{c} \tilde{\phi}_{+} \\ \tilde{\phi}_{-} \end{array}
\right), \label{eq-stat-W-1}
\end{equation}
the square of the quasiparticle weight becomes
\begin{eqnarray}
Z^2 &=& | \langle \tilde{\Omega}' | \Omega' \rangle |^2 = \left|
\det \tilde{W}_{-,-} \right|^2
\nonumber \\
&=& \det \big\{ \tilde{W}_{-,-}^{\phantom{\dag}} \cdot
\tilde{W}_{-,-}^{\dag} \big\}. \label{eq-stat-Z-1}
\end{eqnarray}
Introducing $G = I - \tilde{W}_{-,-}^{\phantom{\dag}} \cdot
\tilde{W}_{-,-}^{\dag}$, where $I$ is the identity matrix, this
determinant can then be written as
\begin{eqnarray}
Z^2 &=& \exp \left[ \mathrm{tr} \left\{ \log \left[ I - G \right]
\right\} \right] = \exp \left[ -\sum_{r=1}^{\infty} \frac{T_r} {r}
\right],
\label{eq-stat-Z-2} \\
T_r &=& \mathrm{tr} \left\{ G^r \right\} = \sum_{m_1, \ldots, m_r}
G_{m_1, m_2} G_{m_2, m_3} \ldots G_{m_r, m_1}. \nonumber
\end{eqnarray}
Importantly, since the matrix $\tilde{W}$ is unitary, a generic
matrix element of $G$ takes the form
\begin{eqnarray}
G_{m, m'} &=& \delta_{m, m'} - \sum_n \tilde{W}_{m, -, n,
-}^{\phantom{*}} \tilde{W}_{m', -, n, -}^{*}
\nonumber \\
&=& \sum_n \tilde{W}_{m, -, n, +}^{\phantom{*}} \tilde{W}_{m', -, n,
+}^{*}. \label{eq-stat-G}
\end{eqnarray}
Furthermore, the (real) eigenvalues of $G$ are all non-negative
because $\tilde{W} \cdot \tilde{W}^{\dag} = I$, and each expansion
term $T_r$ in Eq.~(\ref{eq-stat-Z-2}) is therefore generically
non-negative.

To calculate the expansion terms in Eq.~(\ref{eq-stat-Z-2}), we must
determine the matrix elements of $\tilde{W}$. Writing $\phi_m^{\dag}
= \sum_{\mathbf{r}} \varphi_{m, \mathbf{r}}^{\phantom{\dag}}
f_{\mathbf{r}}^{\dag}$ and $\tilde{\phi}_m^{\dag} =
\sum_{\mathbf{r}} \tilde{\varphi}_{m, \mathbf{r}}^{\phantom{\dag}}
f_{\mathbf{r}}^{\dag}$ with the label $m$ now running over both
positive-energy and negative-energy levels, the single-particle
wavefunctions $\varphi_{m, \mathbf{r}}$ and $\tilde{\varphi}_{m,
\mathbf{r}}$ are related by
\begin{equation}
\tilde{\varphi}_{n, \mathbf{r}} = \sum_m \tilde{W}_{m, n}
\varphi_{m, \mathbf{r}}, \label{eq-stat-varphi-1}
\end{equation}
and they respectively satisfy
\begin{eqnarray}
&& \sum_{\mathbf{r}'} \mathcal{H}_{\mathbf{r}, \mathbf{r}'}
\varphi_{m, \mathbf{r}'} = \epsilon_m \varphi_{m, \mathbf{r}},
\label{eq-stat-varphi-2} \\
&& \sum_{\mathbf{r}'} \tilde{\mathcal{H}}_{\mathbf{r}, \mathbf{r}'}
\tilde{\varphi}_{n, \mathbf{r}'} = \tilde{\epsilon}_n
\tilde{\varphi}_{n, \mathbf{r}}, \label{eq-stat-varphi-3}
\end{eqnarray}
where the single-particle Hamiltonians can be written as
\begin{eqnarray}
\mathcal{H}_{\mathbf{r}, \mathbf{r}'} &=& 2 J_0 \sum_{\alpha}
\delta_{\mathbf{r} \pm \hat{\mathbf{r}}_{\alpha}, \mathbf{r'}},
\label{eq-stat-H-3} \\
\tilde{\mathcal{H}}_{\mathbf{r}, \mathbf{r}'} &=&
\mathcal{H}_{\mathbf{r}, \mathbf{r}'} +
\tilde{\mathcal{V}}_{\mathbf{r}, \mathbf{r}'}, \quad \,\,
\tilde{\mathcal{V}}_{\mathbf{r}, \mathbf{r}'} = \lim_{V \rightarrow
\infty} \big\{ V \delta_{\mathbf{r}, \mathbf{0}} \,
\delta_{\mathbf{r}', \mathbf{0}} \big\}. \nonumber
\end{eqnarray}
Note that the particle-hole symmetry of the perturbed Hamiltonian is
broken by the finite potential $V$ but is restored in the limit of
infinite potential $V \rightarrow \infty$. Substituting
Eq.~(\ref{eq-stat-varphi-1}) into Eq.~(\ref{eq-stat-varphi-3}), and
using Eq.~(\ref{eq-stat-varphi-2}) gives
\begin{equation}
\sum_m \tilde{W}_{m, n} \epsilon_m \varphi_{m, \mathbf{r}} +
\sum_{\mathbf{r}'} \tilde{\mathcal{V}}_{\mathbf{r}, \mathbf{r}'}
\tilde{\varphi}_{n, \mathbf{r}'} = \tilde{\epsilon}_n \sum_m
\tilde{W}_{m, n} \varphi_{m, \mathbf{r}}. \label{eq-stat-varphi-4}
\end{equation}
The matrix element $\tilde{W}_{m, n}$ can then be expressed as
\begin{eqnarray}
\tilde{W}_{m, n} &=& \left( \tilde{\epsilon}_n - \epsilon_m
\right)^{-1} \sum_{\mathbf{r}, \mathbf{r'}}
\tilde{\mathcal{V}}_{\mathbf{r}, \mathbf{r}'}^{\phantom{*}}
\varphi_{m, \mathbf{r}}^{*} \tilde{\varphi}_{n,
\mathbf{r}'}^{\phantom{*}}
\nonumber \\
&=& \frac{V \varphi_{m, \mathbf{0}}^{*} \, \tilde{\varphi}_{n,
\mathbf{0}}^{\phantom{*}}} {\tilde{\epsilon}_n - \epsilon_m} \, .
\label{eq-stat-W-2}
\end{eqnarray}
This result for $\tilde{W}_{m, n}$ is not final because we do not
know the perturbed energies $\tilde{\epsilon}_n$ or even the
perturbed wavefunction $\tilde{\varphi}_{n, \mathbf{0}}$ at site
$\mathbf{0}$. However, since the matrix $\tilde{W}$ is unitary, its
matrix elements satisfy the normalization condition
\begin{equation}
\sum_m | \tilde{W}_{m, n} |^2 = \tilde{\mathcal{N}}_n \sum_m \left(
\tilde{\epsilon}_n - \epsilon_m \right)^{-2} = 1,
\label{eq-stat-norm-1}
\end{equation}
where $\tilde{\mathcal{N}}_n = V^2 |\tilde{\varphi}_{n,
\mathbf{0}}|^2 / N$ is a normalization constant, and $N$ is the
system size. Note that $|\varphi_{m, \mathbf{0}}|^2 = 1/N$ for all
$m$ due to the translation symmetry of the unperturbed system.
Furthermore, substituting Eq.~(\ref{eq-stat-W-2}) into
Eq.~(\ref{eq-stat-varphi-1}), and setting $\mathbf{r} = \mathbf{0}$
results in the self-consistency condition
\begin{equation}
\frac{1}{N} \sum_m \left( \tilde{\epsilon}_n - \epsilon_m
\right)^{-1} = \frac{1}{V} \rightarrow 0. \label{eq-stat-sc-1}
\end{equation}
Note that the opposite limit $V = 0$ corresponds to the unperturbed
system and gives $\tilde{\epsilon}_n = \epsilon_n$ for all levels.

We use Eqs.~(\ref{eq-stat-norm-1}) and (\ref{eq-stat-sc-1}) to
determine the matrix elements $\tilde{W}_{m, n}$ via the
normalization constant $\tilde{\mathcal{N}}_n$ and the perturbed
energies $\tilde{\epsilon}_n$. Since the perturbation
$\tilde{\mathcal{V}}_{\mathbf{r}, \mathbf{r}'}$ is represented by a
rank-$1$ matrix, it couples to only one (suitably chosen) level
within any set of degenerate levels, and the unperturbed energies
$\epsilon_m$ in Eq.~(\ref{eq-stat-sc-1}) are therefore effectively
non-degenerate. Each perturbed energy $\tilde{\epsilon}_n$ satisfies
$\epsilon_n \leq \tilde{\epsilon}_n \leq \epsilon_{n+1}$, and the
sum in $m$ can be turned into a (non-divergent) integral for $m \neq
\{ n, n+1 \}$. However, since $\tilde{\epsilon}_n$ can be
arbitrarily close to either $\epsilon_n$ or $\epsilon_{n+1}$, the
corresponding two terms must be treated separately. Setting the
overall energy scale to $J_0 = 1$ for simplicity, the schematic form
of Eq.~(\ref{eq-stat-sc-1}) is then
\begin{equation}
\mathbb{P} \int_{-1}^1 \frac{d\epsilon \, g (\epsilon)}
{\tilde{\epsilon}_n - \epsilon} + \frac{1}{N} \left( \frac{1}
{\tilde{\epsilon}_n - \epsilon_n} + \frac{1} {\tilde{\epsilon}_n -
\epsilon_{n+1}} \right) = 0, \label{eq-stat-sc-2}
\end{equation}
where $g (\epsilon) \sim |\epsilon|$ is the density of states around
a Dirac point in two dimensions. Close to the Fermi energy, when
$|\tilde{\epsilon}_n| \ll 1$, the integral in
Eq.~(\ref{eq-stat-sc-2}) is approximately
\begin{equation}
\mathbb{P} \int_{-1}^1 \frac{d\epsilon \, |\epsilon|}
{\tilde{\epsilon}_n - \epsilon} \sim -\tilde{\epsilon}_n \log \left(
1 / |\tilde{\epsilon}_n| \right). \label{eq-stat-int}
\end{equation}
Just above (below) the Fermi energy, when $\tilde{\epsilon}_n > 0$
($\tilde{\epsilon}_n < 0$), this integral is negative (positive),
and the perturbed energy $\tilde{\epsilon}_n$ is therefore closest
to $\epsilon_n$ ($\epsilon_{n+1}$) among the unperturbed energies
$\epsilon_m$. In either case, the corresponding minimal energy
difference is $\min_m |\tilde{\epsilon}_n - \epsilon_m| \sim [N
|\tilde{\epsilon}_n| \log (1 / |\tilde{\epsilon}_n|)]^{-1}$, which
is parametrically smaller than the mean level spacing $[N g
(\tilde{\epsilon}_n)]^{-1} \sim [N |\tilde{\epsilon}_n|]^{-1}$ at
the given energy. The sum in Eq.~(\ref{eq-stat-norm-1}) is then
dominated by this minimal energy difference, and the normalization
constant becomes
\begin{equation}
\tilde{\mathcal{N}}_n \sim \min_m \left( \tilde{\epsilon}_n -
\epsilon_m \right)^2 \sim \left[ N \tilde{\epsilon}_n \log (1 /
|\tilde{\epsilon}_n|) \right]^{-2}. \label{eq-stat-norm-2}
\end{equation}
Substituting Eq.~(\ref{eq-stat-norm-2}) into
Eq.~(\ref{eq-stat-W-2}), the absolute value of the matrix element
$\tilde{W}_{m, n}$ takes the form
\begin{equation}
|\tilde{W}_{m, n}| \sim \frac{\left[ N |\tilde{\epsilon}_n| \log (1
/ |\tilde{\epsilon}_n|) \right]^{-1}} {|\tilde{\epsilon}_n -
\epsilon_m|} \, . \label{eq-stat-W-3}
\end{equation}
The matrix element itself could in principle have a complex phase
factor, but it is not necessary as $\varphi_{m, \mathbf{0}}$ and
$\tilde{\varphi}_{n, \mathbf{0}}$ can all be set real
simultaneously.

We are now ready to calculate the quasiparticle weight via the
expansion terms $T_r$ in Eq.~(\ref{eq-stat-Z-2}). In particular, by
using Eq.~(\ref{eq-stat-G}), the first expansion term becomes
\begin{eqnarray}
T_1 &=& \sum_m G_{m, m} = \sum_{m, n} \left| \tilde{W}_{m, -, n, +}
\right|^2 \nonumber \\
&\sim& \sum_{\tilde{\epsilon}_n > 0} \, \sum_{\epsilon_m < 0}
\frac{[N \tilde{\epsilon}_n \log (1 / |\tilde{\epsilon}_n|)]^{-2}}
{(\tilde{\epsilon}_n - \epsilon_m)^2} \, . \label{eq-stat-T-1}
\end{eqnarray}
Turning the sums into integrals and using $g (\epsilon) \sim
|\epsilon|$, this expansion term takes the schematic low-energy form
\begin{eqnarray}
T_1 &\sim& \int_0^1 d\tilde{\epsilon} \int_0^1 d\epsilon \, \frac{g
(\tilde{\epsilon}) g (\epsilon)} {(\tilde{\epsilon} + \epsilon)^2
[\tilde{\epsilon} \log (1 / \tilde{\epsilon})]^2}
\nonumber \\
&\sim& \int_0^1 \frac{d\tilde{\epsilon}} {\tilde{\epsilon} \log (1 /
\tilde{\epsilon})} \, . \label{eq-stat-T-2}
\end{eqnarray}
The infrared divergence at $\tilde{\epsilon} = 0$ can be regularized
by a cutoff at $\tilde{\epsilon} \sim 1/N$ for any finite system
size $N$. The first expansion term is then $T_1 \sim \log \log N$.
Since the remaining expansion terms $T_r$ are all non-negative, the
square of the quasiparticle weight can be bounded from above as
\begin{equation}
Z^2 \leq \exp \left[ -T_1 \right] \sim \exp \left[ -\kappa \log \log
N \right] \sim \left[ \log N \right]^{-\kappa}, \label{eq-stat-Z-3}
\end{equation}
where $\kappa$ is an unknown positive exponent. Regardless of its
precise value, the quasiparticle weight $Z$ vanishes in the limit of
$N \rightarrow \infty$. We therefore conclude that there is an
orthogonality catastrophe in the stationary limit. Note that this
orthogonality catastrophe is weaker than in the standard case
because the ground-state overlap decays with a logarithm and not
with a power law of the system size.\cite{Anderson-3} This
difference is explained by the smaller density of states around the
Fermi energy, which is linear in our case and not constant as in the
standard case. In fact, for a finite potential $V$ in
Eq.~(\ref{eq-stat-H-3}), there would no longer be an orthogonality
catastrophe.\cite{Hentschel}

\section{Variational approach} \label{sec-var}

If the hole is mobile ($t > 0$), the doped Hamiltonian $\tilde{H}_c
(\mathbf{K})$ in Eq.~(\ref{eq-gen-H-7}) is no longer quadratic, and
therefore the ground state of the doped model is not exactly known.
Furthermore, unlike in the gapped phase of the model,\cite{Halasz-1}
there is no well-controlled perturbative approach around the
stationary limit because there are infinitely many low-energy
eigenstates arbitrarily close to the stationary ground state. To
obtain an approximate ground state for a slow hole ($t \ll J_0$), we
employ a variational approach in terms of a single-parameter trial
state that interpolates smoothly between the ground state for a
stationary hole and that for an infinitely fast hole.

\begin{figure}[h]
\centering
\includegraphics[width=0.85\columnwidth]{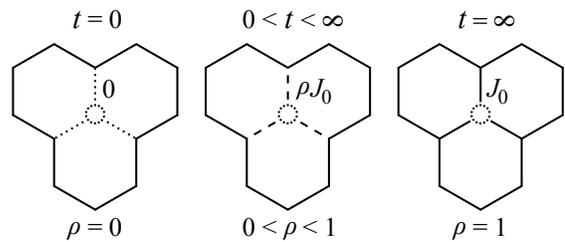}
\caption{Fermion coupling strengths around the instantaneous hole
site (dotted circle) for the quadratic variational Hamiltonian
$\hat{H}_c (\varrho)$ at $t = 0$ (left), finite $t$ (middle), and $t
\rightarrow \infty$ (right). Default couplings $J_0$ are marked by
solid lines, partially switched-off couplings $\varrho J_0$ are
marked by dashed lines (where $0 < \varrho < 1$), and fully
switched-off couplings $0$ are marked by dotted lines.}
\label{fig-2}
\end{figure}

In the limit of a stationary hole ($t = 0$), the doped ground state
is known exactly (see Sec.~\ref{sec-stat}). In the doped
Hamiltonian, the fermion coupling terms of the undoped Hamiltonian
are fully switched off around the hole site [see
Eq.~(\ref{eq-gen-H-7})]. In the limit of an infinitely fast hole ($t
\rightarrow \infty$), we expect that the doped ground state is close
to the undoped ground state because the hole can hop most freely in
its translation-symmetric fermion configuration. In particular, the
expectation value of the inversion operator $\hat{R}_{\alpha}$ is
maximized in Eq.~(\ref{eq-gen-H-7}) by the undoped ground state as
$\langle \omega | \hat{R}_{\alpha} | \omega \rangle = +1$. For the
trial state at $t \ll J_0$, it is then natural to choose the ground
state $| \hat{\omega} \rangle$ of the Hamiltonian $\hat{H}_c
(\varrho) = (1 - \varrho) \tilde{H}_c + \varrho H_c$, which
interpolates smoothly between the stationary doped Hamiltonian
$\tilde{H}_c$ and the undoped Hamiltonian $H_c$ as a function of a
variational parameter $0 \leq \varrho \leq 1$. In this Hamiltonian,
the fermion coupling terms of the undoped Hamiltonian are partially
switched off around the instantaneous hole site (see
Fig.~\ref{fig-2}).

Note that our intuition for the infinitely fast hole appears to be
in contradiction with Nagaoka's theorem\cite{Nagaoka} which predicts
a spin-polarized ground state for $t \rightarrow \infty$. The reason
for this apparent contradiction is that the flux excitations are not
negligible for $t \gg J_0$. From the point of view of such a
high-energy hole, the spins are not fractionalized into fluxes and
fermions. In this regime, the translation-symmetric spin
configuration of Nagaoka's ground state minimizes the kinetic energy
of the hole by maximizing its effective hopping amplitude between
neighboring hole positions. However, it is possible to imagine a
scenario in which $E_{\mathrm{fermion}}^{\max} \ll t \ll
E_{\mathrm{flux}}^{\min}$ and the spins are fractionalized into
(high-energy) fluxes and (low-energy) fermions from the point of
view of the hole. In this regime, the translation-symmetric fermion
configuration of the undoped ground state minimizes the kinetic
energy within the zero-flux sector. In some sense, it is the natural
generalization of Nagaoka's ground state for this fractionalized
scenario. Although there is no such intermediate regime for the
Kitaev model due to $E_{\mathrm{fermion}}^{\max} \sim
E_{\mathrm{flux}}^{\min} \sim J_0$, the fluxes are nevertheless
negligible for $t \ll J_0$, and it is therefore reasonable to choose
a trial state that interpolates between the stationary doped and the
undoped ground states.

Since the Hamiltonian $\hat{H}_c (\varrho)$ is quadratic, the
quasiparticle weight $Z = | \langle \hat{\omega} | \omega \rangle
|^2$ in terms of the trial state $| \hat{\omega} \rangle$ can be
calculated in exactly the same way as in Sec.~\ref{sec-stat}. Using
the doubled formulation, the perturbed fermions $\hat{\phi}_m$ and
the unperturbed fermions $\phi_m$ are related to each other by
Eq.~(\ref{eq-stat-W-1}), but via a different unitary matrix
$\hat{W}$. The corresponding single-particle wavefunctions are then
related by
\begin{equation}
\hat{\varphi}_{n, \mathbf{r}} = \sum_m \hat{W}_{m, n} \varphi_{m,
\mathbf{r}}, \label{eq-var-varphi-1}
\end{equation}
and the perturbed wavefunction satisfies
\begin{equation}
\sum_{\mathbf{r}'} \hat{\mathcal{H}}_{\mathbf{r}, \mathbf{r}'}
\hat{\varphi}_{n, \mathbf{r}'} = \hat{\epsilon}_n \hat{\varphi}_{n,
\mathbf{r}}, \label{eq-var-varphi-2}
\end{equation}
where the appropriate single-particle Hamiltonian is
\begin{eqnarray}
\hat{\mathcal{H}}_{\mathbf{r}, \mathbf{r}'} &=&
\mathcal{H}_{\mathbf{r}, \mathbf{r}'} +
\hat{\mathcal{V}}_{\mathbf{r}, \mathbf{r}'},
\label{eq-var-H} \\
\hat{\mathcal{V}}_{\mathbf{r}, \mathbf{r}'} &=& -(1 - \varrho) \,
\mathcal{H}_{\mathbf{r}, \mathbf{r}'} \left( \delta_{\mathbf{r},
\mathbf{0}} + \delta_{\mathbf{r}', \mathbf{0}} \right). \nonumber
\end{eqnarray}
Using Eq.~(\ref{eq-stat-W-2}), the matrix element $\hat{W}_{m, n}$
is then given by
\begin{eqnarray}
\hat{W}_{m, n} &=& \left( \hat{\epsilon}_n - \epsilon_m \right)^{-1}
\sum_{\mathbf{r}, \mathbf{r'}} \hat{\mathcal{V}}_{\mathbf{r},
\mathbf{r}'}^{\phantom{*}} \varphi_{m, \mathbf{r}}^{*}
\hat{\varphi}_{n, \mathbf{r}'}^{\phantom{*}}
\label{eq-var-W-1} \\
&=& -(1 - \varrho) \, \frac{\epsilon_m \varphi_{m, \mathbf{0}}^{*}
\hat{\varphi}_{n, \mathbf{0}}^{\phantom{*}} + \varphi_{m,
\mathbf{0}}^{*} \hat{\varphi}_{n, \mathbf{0}}'} {\hat{\epsilon}_n -
\epsilon_m} \, , \nonumber
\end{eqnarray}
where $\hat{\varphi}_{n, \mathbf{0}}' \equiv \sum_{\mathbf{r}}
\mathcal{H}_{\mathbf{0}, \mathbf{r}}^{\phantom{*}} \hat{\varphi}_{n,
\mathbf{r}}^{\phantom{*}}$. Since the matrix $\hat{W}$ is unitary,
its matrix elements satisfy the normalization condition
\begin{equation}
\sum_m | \hat{W}_{m, n} |^2 = \hat{\mathcal{N}}_n \sum_m \left|
\hat{\xi}_n + \epsilon_m \right|^2 \left( \hat{\epsilon}_n -
\epsilon_m \right)^{-2} = 1, \label{eq-var-norm-1}
\end{equation}
where $\hat{\mathcal{N}}_n = (1 - \varrho)^2 |\hat{\varphi}_{n,
\mathbf{0}}|^2 / N$ is a normalization constant, and $\hat{\xi}_n
\equiv \hat{\varphi}_{n, \mathbf{0}}' / \hat{\varphi}_{n,
\mathbf{0}}^{\phantom{*}}$ is a wavefunction ratio. Note again that
$|\varphi_{m, \mathbf{0}}|^2 = 1/N$ for all $m$ due to the
translation symmetry of the unperturbed system. Using
Eqs.~(\ref{eq-stat-varphi-2}), (\ref{eq-var-varphi-1}), and
(\ref{eq-var-W-1}), we also obtain two independent self-consistency
conditions
\begin{eqnarray}
\hat{\varphi}_{n, \mathbf{0}}^{\phantom{*}} &=& \sum_m \hat{W}_{m,
n} \varphi_{m, \mathbf{0}} = -\frac{1 - \varrho} {N} \, \sum_m
\frac{\epsilon_m \hat{\varphi}_{n, \mathbf{0}}^{\phantom{*}} +
\hat{\varphi}_{n, \mathbf{0}}'} {\hat{\epsilon}_n - \epsilon_m} \, ,
\nonumber \\
\hat{\varphi}_{n, \mathbf{0}}' &=& \sum_m \hat{W}_{m, n}
\sum_{\mathbf{r}} \mathcal{H}_{\mathbf{0}, \mathbf{r}} \varphi_{m,
\mathbf{r}} = \sum_m \hat{W}_{m, n} \epsilon_m \varphi_{m,
\mathbf{0}} \nonumber \\
&=& -\frac{1 - \varrho} {N} \, \sum_m \frac{\epsilon_m^2
\hat{\varphi}_{n, \mathbf{0}}^{\phantom{*}} + \epsilon_m
\hat{\varphi}_{n, \mathbf{0}}'} {\hat{\epsilon}_n - \epsilon_m} \, .
\label{eq-var-sc-1}
\end{eqnarray}
Demanding non-trivial solutions for $\hat{\varphi}_{n,
\mathbf{0}}^{\phantom{*}}$ and $\hat{\varphi}_{n, \mathbf{0}}'$
leads to the combined self-consistency condition
\begin{equation}
\left| \begin{array}{cc} 1 + (1 - \varrho) \Sigma_n' & (1 - \varrho)
\Sigma_n \\ (1 - \varrho) \Sigma_n'' & 1 + (1 - \varrho) \Sigma_n'
\end{array} \right| = 0, \label{eq-var-sc-2}
\end{equation}
where the sums $\Sigma_n$, $\Sigma_n'$, and $\Sigma_n''$ are given
by
\begin{eqnarray}
\Sigma_n &=& \frac{1}{N} \sum_m \left( \hat{\epsilon}_n - \epsilon_m
\right)^{-1},
\nonumber \\
\Sigma_n' &=& \frac{1}{N} \sum_m \epsilon_m \left( \hat{\epsilon}_n
- \epsilon_m \right)^{-1} = \hat{\epsilon}_n \Sigma_n - 1,
\label{eq-var-sigma} \\
\Sigma_n'' &=& \frac{1}{N} \sum_m \epsilon_m^2 \left(
\hat{\epsilon}_n - \epsilon_m \right)^{-1} = \hat{\epsilon}_n^2
\Sigma_n - \hat{\epsilon}_n. \nonumber
\end{eqnarray}
Note that $\sum_m 1 = N$ counts the number of energy levels and that
$\sum_m \epsilon_m = 0$ due to particle-hole symmetry. Substituting
Eq.~(\ref{eq-var-sigma}) into Eq.~(\ref{eq-var-sc-2}), the
self-consistency condition becomes
\begin{equation}
\Sigma_n = \frac{1}{N} \sum_m \left( \hat{\epsilon}_n - \epsilon_m
\right)^{-1} = -\frac{\varrho^2} {(1 - \varrho^2) \hat{\epsilon}_n}
\, . \label{eq-var-sc-3}
\end{equation}
This result reduces to the $V \rightarrow \infty$ limit of
Eq.~(\ref{eq-stat-sc-1}) for $\varrho = 0$ and the $V = 0$ limit of
Eq.~(\ref{eq-stat-sc-1}) for $\varrho = 1$. However,
Eq.~(\ref{eq-var-sc-3}) is particle-hole symmetric for all values of
$0 \leq \varrho \leq 1$ as it is invariant under $\epsilon_m
\rightarrow -\epsilon_m$ and $\hat{\epsilon}_n \rightarrow
-\hat{\epsilon}_n$.

We use Eqs.~(\ref{eq-var-norm-1}), (\ref{eq-var-sc-1}), and
(\ref{eq-var-sc-3}) to determine the matrix elements $\hat{W}_{m,
n}$ via the normalization constant $\hat{\mathcal{N}}_n$, the
wavefunction ratio $\hat{\xi}_n$, and the perturbed energies
$\hat{\epsilon}_n$. Although the perturbation
$\hat{\mathcal{V}}_{\mathbf{r}, \mathbf{r}'}$ is represented by a
rank-$2$ matrix, it only couples to levels that have finite
wavefunctions at site $\mathbf{0}$. Since there is only one such
(suitably chosen) level within any set of degenerate levels, the
unperturbed energies $\epsilon_m$ in Eq.~(\ref{eq-var-sc-3}) are
still effectively non-degenerate. Turning the sum in $m$ into an
integral, but treating the terms with $m = \{ n, n+1 \}$ separately,
the schematic form of Eq.~(\ref{eq-var-sc-3}) is
\begin{equation}
\frac{1}{N} \left( \frac{1} {\hat{\epsilon}_n - \epsilon_n} +
\frac{1} {\hat{\epsilon}_n - \epsilon_{n+1}} \right) =
-\frac{\varrho^2} {(1 - \varrho^2) \hat{\epsilon}_n} \, ,
\label{eq-var-sc-4}
\end{equation}
where the integral is immediately neglected because its value $\sim
\hat{\epsilon}_n \log (1 / |\hat{\epsilon}_n|)$ [see
Eq.~(\ref{eq-stat-int})] is much smaller than the term $\sim 1 /
\hat{\epsilon}_n$ for any $\varrho > 0$ and $|\hat{\epsilon}_n| \ll
1$. For both $\hat{\epsilon}_n > 0$ and $\hat{\epsilon}_n < 0$, the
sum in Eq.~(\ref{eq-var-norm-1}) is then dominated by the minimal
energy difference $\min_m |\hat{\epsilon}_n - \epsilon_m| \sim (1 -
\varrho^2) |\hat{\epsilon}_n| / (N \varrho^2)$. Since
Eqs.~(\ref{eq-var-sc-1}) and (\ref{eq-var-sc-3}) give $\hat{\xi}_n =
\hat{\epsilon}_n / \varrho$ for the wavefunction ratio, the
normalization constant becomes
\begin{eqnarray}
\hat{\mathcal{N}}_n &\sim& \min_m \left[ \left| \hat{\xi}_n +
\epsilon_m \right|^{-2} \left( \hat{\epsilon}_n - \epsilon_m
\right)^2 \right]
\label{eq-var-norm-2} \\
&\sim& \frac{(1 - \varrho^2)^2 \hat{\epsilon}_n^2} {N^2 \varrho^4 |
\hat{\xi}_n + \hat{\epsilon}_n |^2} \sim \frac{(1 - \varrho)^2} {N^2
\varrho^2} \, . \nonumber
\end{eqnarray}
Substituting Eq.~(\ref{eq-var-norm-2}) into Eq.~(\ref{eq-var-W-1}),
the absolute value of the matrix element $\hat{W}_{m, n}$ takes the
form
\begin{equation}
|\hat{W}_{m, n}| \sim \frac{(1 - \varrho) \, |\hat{\epsilon}_n +
\varrho \, \epsilon_m|} {N \varrho^2 \, |\hat{\epsilon}_n -
\epsilon_m|} \, . \label{eq-var-W-2}
\end{equation}
The matrix elements again do not need to have complex phase factors
as $\varphi_{m, \mathbf{0}}$ and $\hat{\varphi}_{n, \mathbf{0}}$ can
all be set real simultaneously.

We finally calculate the quasiparticle weight via the expansion
terms $T_r$ in Eq.~(\ref{eq-stat-Z-2}). The first expansion term
reads
\begin{eqnarray}
T_1 &=& \sum_m G_{m, m} = \sum_{m, n} \left| \hat{W}_{m, -, n, +}
\right|^2 \nonumber \\
&\sim& \sum_{\hat{\epsilon}_n > 0} \, \sum_{\epsilon_m < 0} \frac{(1
- \varrho)^2 (\hat{\epsilon}_n + \varrho \, \epsilon_m)^2} {N^2
\varrho^4 (\hat{\epsilon}_n - \epsilon_m)^2} \, . \label{eq-var-T-1}
\end{eqnarray}
Turning the sums into integrals and using $g (\epsilon) \sim
|\epsilon|$, this expansion term takes the schematic low-energy form
\begin{equation}
T_1 \sim \frac{(1 - \varrho)^2} {\varrho^4} \int_0^1 d\hat{\epsilon}
\int_0^1 d\epsilon \, \frac{\hat{\epsilon} \, \epsilon \,
[\hat{\epsilon} - \varrho \, \epsilon]^2} {(\hat{\epsilon} +
\epsilon)^2} \, . \label{eq-var-T-2}
\end{equation}
By counting the powers of $\epsilon$ and $\hat{\epsilon}$, we deduce
that this integral has no infrared divergence. Furthermore, the same
power-counting argument reveals that the integrals for the remaining
expansion terms $T_r$ are also finite. We conclude that there is no
orthogonality catastrophe for any $\varrho > 0$ within the
variational framework and that the quasiparticle weight $Z$ remains
finite in the thermodynamic limit.

To relate the variational results to our original formulation, we
would in principle need to perform a variational optimization that
determines the best possible trial state $| \hat{\omega} \rangle$
for a given hopping amplitude $t$. Such a calculation would give the
best variational parameter $\varrho (t)$ as a function of $t$, which
could then be substituted directly into our variational results.
However, it would require a more accurate calculation of the
quasiparticle weight $Z$ and is beyond the scope of this
work.\cite{Halasz-2} Nevertheless, we would necessarily find
$\varrho > 0$ and hence $Z > 0$ for any $t > 0$. According to
Eq.~(\ref{eq-gen-H-7}), a small hole momentum $\mathbf{K}$ does not
matter either because it only renormalizes the hopping amplitude
along an $\alpha$ bond as $t \rightarrow t \cos [\mathbf{K} \cdot
\hat{\mathbf{r}}_{\alpha} + \tilde{\vartheta}_{\mathbf{K}}]$. We
therefore anticipate that the hole propagates as a coherent
quasiparticle for any hopping amplitude $t > 0$ and any hole
momentum $|\mathbf{K}| \ll | \hat{\mathbf{r}}_{\alpha} |^{-1}$. The
quasiparticle weight is then finite for $t > 0$ but vanishes in the
limit of $t \rightarrow 0$.

\section{One-dimensional limit} \label{sec-1D}

As a complementary direction to the variational approach, we
consider a spatially anisotropic special point in the gapless phase
characterized by $J_{x,y} = J_0$ and $J_z = 0$, where the Kitaev
model breaks down into non-interacting one-dimensional (1D) chains
along the $x$ and $y$ bonds. Exploiting the relative simplicity of
this 1D limit, and studying a modified problem that is
asymptotically (i.e., for the lowest-energy fermions) equivalent to
the original one, we can then determine if there is an orthogonality
catastrophe for a mobile hole ($t > 0$) without resorting to a
variational framework.

\begin{figure}[h]
\centering
\includegraphics[width=0.7\columnwidth]{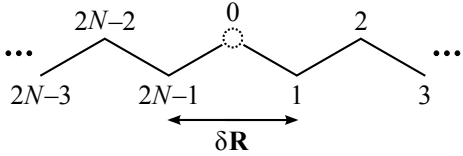}
\caption{Illustration of the one-dimensional (1D) chain with the
lattice constant $\delta \mathbf{R}$ and the site labeling
convention around the instantaneous hole site $\ell = 0$ (dotted
circle).} \label{fig-3}
\end{figure}

For a single 1D chain of length $2N$, the sites are labeled by $\ell
= \{ 1, 2, \ldots, 2N \equiv 0 \}$, and the instantaneous hole site
$\mathbf{0} \in A$ is labeled by $\ell = 0$ (see Fig.~\ref{fig-3}).
The hole momentum is $K = \mathbf{K} \cdot \delta \mathbf{R}$, and
the fermion momenta are $k = \mathbf{k} \cdot \delta \mathbf{R}$,
where $\delta \mathbf{R} = \hat{\mathbf{r}}_y - \hat{\mathbf{r}}_x$
is the lattice constant. Since the even ($\eta$) and odd ($\mu$)
fermions $\psi_{k, \eta} (\alpha)$ and $\psi_{k, \mu} (\alpha)$ that
diagonalize the inversion operator $\hat{R}_{\alpha}$ in
Eq.~(\ref{eq-gen-R}) are defined for pairs of momenta $\pm k$ [see
Eq.~(\ref{eq-gen-psi-1})], we restrict our attention to non-negative
momenta $k = \{ 0, \delta k, 2 \delta k, \ldots, \pi \}$, where
$\delta k = 2\pi / N$ is the momentum spacing. The total number of
complex fermions is then $N$ because there are two fermions
$\psi_{k, \eta} (\alpha)$ and $\psi_{k, \mu} (\alpha)$ for each $0 <
k < \pi$ and there is one fermion $\psi_{k, \eta} (\alpha)$ for each
of $k = 0$ and $k = \pi$.

Since the distinction between the two sublattices is entirely
artificial in the 1D chain, the phase difference
$\tilde{\vartheta}_{\mathbf{K}}$ in Eq.~(\ref{eq-gen-H-7}) vanishes
for any hole momentum $K$. Using the 1D notation, the undoped
Hamiltonian in Eq.~(\ref{eq-gen-H-5}) is then
\begin{equation}
H_c = J_0 \sum_{\ell \in A} \left( i c_{\ell} c_{\ell + 1} + i
c_{\ell} c_{\ell - 1} \right), \label{eq-1D-H-1}
\end{equation}
while the doped Hamiltonian in Eq.~(\ref{eq-gen-H-7}) takes the form
\begin{eqnarray}
\tilde{H}_c (K) &=& H_c - J_0 \left( i c_0 c_1 + i c_0 c_{2N-1}
\right) \label{eq-1D-H-2} \\
&& \, -\frac{t}{2} \cos \frac{K}{2} \sum_{\alpha = x,y} \left[
\hat{R}_{\alpha} - i c_0 \hat{R}_{\alpha} c_0 \right]. \nonumber
\end{eqnarray}
The inversion operators $\hat{R}_{x,y}$ are diagonalized by the
fermions $\psi_{k, \eta} (x,y)$ and $\psi_{k, \mu} (x,y)$. In the 1D
notation, the Majorana fermion components of these fermions in
Eq.~(\ref{eq-gen-gamma-1}) are
\begin{eqnarray}
\gamma_{k, \eta, \Xi} (x,y) &\propto& \sum_{\ell \in \Xi} \cos
\left[ \frac{k \ell} {2} \pm \frac{k} {4} \right] c_{\ell},
\nonumber \\
\gamma_{k, \mu, \Xi} (x,y) &\propto& \sum_{\ell \in \Xi} \sin \left[
\frac{k \ell} {2} \pm \frac{k} {4} \right] c_{\ell}.
\label{eq-1D-gamma-1}
\end{eqnarray}
The fermions corresponding to $\hat{R}_x$ and $\hat{R}_y$ also each
diagonalize the undoped Hamiltonian $H_c$ with degenerate energies
$\epsilon_{k, \eta} = \epsilon_{k, \mu} = 4 J_0 \cos (k/2)$.

In the limit of the lowest energies at $k \rightarrow \pi$, the
Majorana fermion components in Eq.~(\ref{eq-1D-gamma-1}) are related
to each other by $\gamma_{k, \eta, \Xi} (y) = \gamma_{k, \mu, \Xi}
(x)$ and $\gamma_{k, \mu, \Xi} (y) = -\gamma_{k, \eta, \Xi} (x)$.
The inversion operators $\hat{R}_x$ and $\hat{R}_y$ are therefore
diagonalized by the same fermions at the lowest energies. This
property motivates us to define modified operators $\bar{R}_x$ and
$\bar{R}_y$ that are diagonalized by the same fermions at all
energies and reduce to the inversion operators $\hat{R}_x$ and
$\hat{R}_y$ at the lowest energies. These modified operators are
given by Eq.~(\ref{eq-gen-R}) but in terms of modified fermions
$\bar{\psi}_{k, \eta} (x,y)$ and $\bar{\psi}_{k, \mu} (x,y)$ that
have modified Majorana fermion components
\begin{eqnarray}
\bar{\gamma}_{k, \eta, \Xi} (x,y) &\propto& \sum_{\ell \in \Xi} \cos
\left[ \frac{k \ell} {2} \pm \frac{\pi} {4} \right] c_{\ell},
\nonumber \\
\bar{\gamma}_{k, \mu, \Xi} (x,y) &\propto& \sum_{\ell \in \Xi} \sin
\left[ \frac{k \ell} {2} \pm \frac{\pi} {4} \right] c_{\ell}.
\label{eq-1D-gamma-2}
\end{eqnarray}
Indeed, $\bar{\gamma}_{k, \eta, \Xi} (x,y)$ and $\bar{\gamma}_{k,
\mu, \Xi} (x,y)$ reduce to $\gamma_{k, \eta, \Xi} (x,y)$ and
$\gamma_{k, \mu, \Xi} (x,y)$ in the limit of $k \rightarrow \pi$,
and they also satisfy
\begin{eqnarray}
\bar{\gamma}_{k, \eta, \Xi} &\equiv& \bar{\gamma}_{k, \eta, \Xi} (x)
= -\bar{\gamma}_{k, \mu, \Xi} (y),
\nonumber \\
\bar{\gamma}_{k, \mu, \Xi} &\equiv& \bar{\gamma}_{k, \mu, \Xi} (x) =
\bar{\gamma}_{k, \eta, \Xi} (y) \label{eq-1D-gamma-3}
\end{eqnarray}
for all $0 \leq k \leq \pi$. Furthermore, the modified fermions
$\bar{\psi}_{k, \eta} \equiv \bar{\psi}_{k, \eta} (x) =
-\bar{\psi}_{k, \mu} (y)$ and $\bar{\psi}_{k, \mu} \equiv
\bar{\psi}_{k, \mu} (x) = \bar{\psi}_{k, \eta} (y)$ still
diagonalize the undoped Hamiltonian $H_c$ with energies $\epsilon_k
= \epsilon_{k, \eta} = \epsilon_{k, \mu} = 4 J_0 \cos (k/2)$.

For the corresponding modified problem, the inversion operators
$\hat{R}_x$ and $\hat{R}_y$ in Eq.~(\ref{eq-1D-H-2}) are replaced by
the modified operators $\bar{R}_x$ and $\bar{R}_y$. On the one hand,
since the orthogonality catastrophe is determined by the
lowest-energy fermions, the modified problem must have the same kind
of orthogonality catastrophe as the original one. On the other hand,
the doped Hamiltonian in the modified problem is simplified
considerably with respect to Eq.~(\ref{eq-1D-H-2}). In particular,
$\bar{R}_x$ and $\bar{R}_y$ are diagonalized by the same fermions,
but each excited fermion multiplies them by opposite factors $\pm
i$. They therefore take identical values for even fermion number and
opposite values for odd fermion number. Since the fermion number is
even for physical states, and the fermion parity is flipped by
$c_0$, several terms in Eq.~(\ref{eq-1D-H-2}) can be related to each
other as
\begin{eqnarray}
\bar{R} &\equiv& \bar{R}_x = \bar{R}_y,
\label{eq-1D-R} \\
i c_0 \bar{R} c_0 &=& i c_0 \bar{R}_x c_0 = -i c_0 \bar{R}_y c_0.
\nonumber
\end{eqnarray}
In terms of the modified Majorana fermion components
$\bar{\gamma}_{k, \eta, \Xi}$ and $\bar{\gamma}_{k, \mu, \Xi}$, the
undoped Hamiltonian in Eq.~(\ref{eq-1D-H-1}) is then
\begin{eqnarray}
H_c &=& \frac{i}{2} \sum_k \epsilon_k \left( \bar{\gamma}_{k, \eta,
A} \, \bar{\gamma}_{k, \eta, B} + \bar{\gamma}_{k, \mu, A} \,
\bar{\gamma}_{k, \mu, B} \right)
\nonumber \\
&=& \sum_k \epsilon_k \left( \bar{\psi}_{k, \eta}^{\dag}
\bar{\psi}_{k, \eta}^{\phantom{\dag}} + \bar{\psi}_{k, \mu}^{\dag}
\bar{\psi}_{k, \mu}^{\phantom{\dag}} - 1 \right), \label{eq-1D-H-3}
\end{eqnarray}
while the doped Hamiltonian in Eq.~(\ref{eq-1D-H-2}) takes the form
\begin{eqnarray}
\bar{H}_c (K) &=& H_c - \frac{i}{2N} \sum_{k, k'} \Big[
\epsilon_{k'} \left( \bar{\gamma}_{k, \eta, A} + \bar{\gamma}_{k,
\mu, A} \right)
\nonumber \\
&& \, \times \left( \bar{\gamma}_{k', \eta, B} + \bar{\gamma}_{k',
\mu, B} \right) \Big] - t_K \bar{R}. \label{eq-1D-H-4}
\end{eqnarray}
Note that this Hamiltonian only depends on the hopping amplitude $t$
and the hole momentum $K$ via the renormalized hopping amplitude
$t_K = t \cos (K/2)$.

The modified problem characterized by Eqs.~(\ref{eq-1D-H-3}) and
(\ref{eq-1D-H-4}) has two important properties. First, like the
original problem in Eqs.~(\ref{eq-1D-H-1}) and (\ref{eq-1D-H-2}), it
has a single energy scale $t_K$ at energies much less than $J_0$.
The corresponding spectral function therefore must take the
universal functional form
\begin{equation}
\mathcal{A} (\varepsilon, K) = \mathcal{N} \left[ \varepsilon -
\bar{E}_{K,0} \right]^{-\zeta} F \left( \frac{\varepsilon -
\bar{E}_{K,0}} {t_K} \right) \label{eq-1D-A}
\end{equation}
in the energy range $0 < \varepsilon - \bar{E}_{K,0} \ll J_0$, where
$\bar{E}_{K,0}$ is the ground-state energy of $\bar{H}_c (K)$. The
function $F$ and the exponent $\zeta$ are universal but unknown,
while the normalization constant $\mathcal{N}$ and the ground-state
energy $\bar{E}_{K,0}$ depend on both $t_K$ and $J_0$. Nevertheless,
the spectral function has the same low-energy functional form for
all $t_K > 0$, up to a rescaling with $t_K$, a renormalization, and
a constant shift in $\varepsilon$, while its low-energy functional
form is a fully self-similar power law for $t_K = 0$. Second, the
Hamiltonian $\bar{H}_c (K)$ is effectively quadratic for both $t_K =
0$ and $t_K \rightarrow \infty$, and its ground state $|
\bar{\omega}_K^{\phantom{\dag}} \rangle$ is therefore known exactly
in both of these limits. Since the Hamiltonian $H_c$ is also
quadratic, the quasiparticle weight $Z = | \langle
\bar{\omega}_K^{\phantom{\dag}} | \omega \rangle |^2$ can then be
calculated exactly.

\begin{figure}[h]
\centering
\includegraphics[width=1.0\columnwidth]{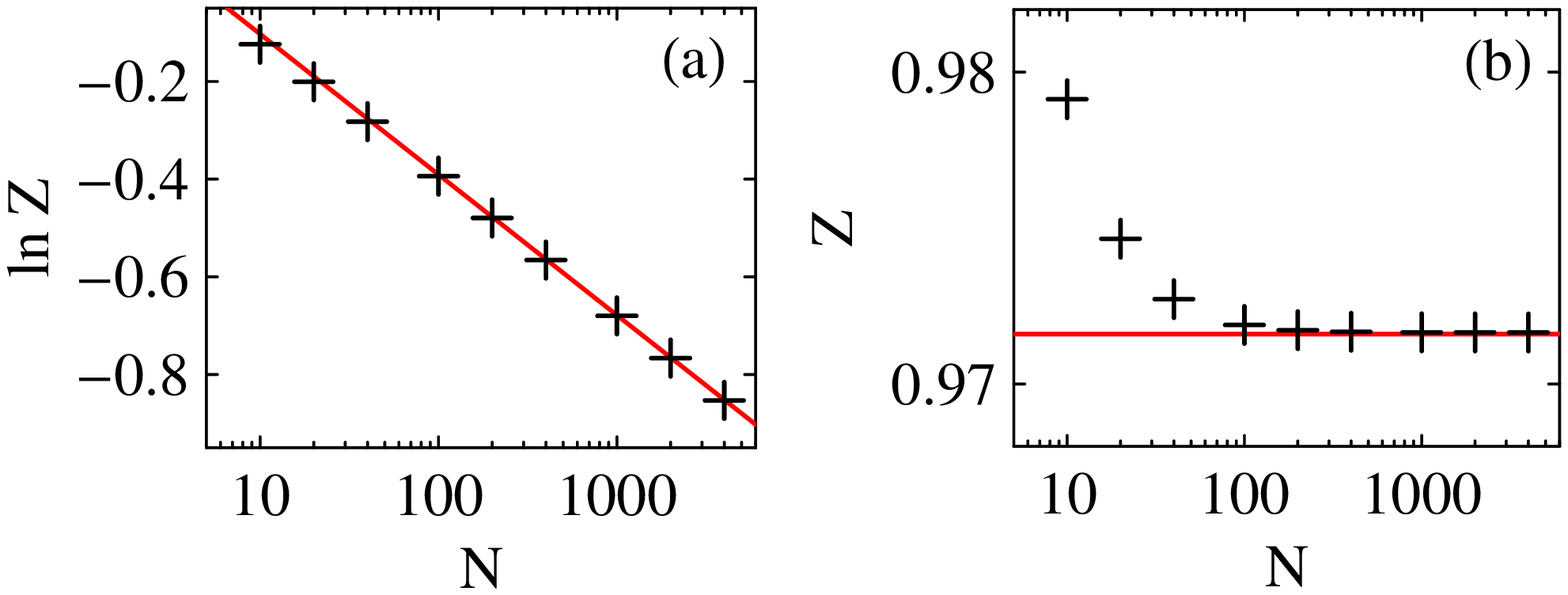}
\caption{Quasiparticle weight $Z$ as a function of the system size
$N$ for $t_K = 0$ (a) and for $t_K \rightarrow \infty$ (b).
Numerical data points are marked by black crosses, while asymptotic
fits of the form $Z \propto N^{-0.125}$ (a) and $Z \approx 0.9716$
(b) are marked by red lines.} \label{fig-4}
\end{figure}

For both $t_K = 0$ and $t_K \rightarrow \infty$, we calculate the
quasiparticle weight numerically for up to $N = 4000$, and study its
behavior in the range $10 \leq N \leq 4000$ (see Fig.~\ref{fig-4}).
For $t_K = 0$, Eq.~(\ref{eq-1D-H-4}) is obviously quadratic, and the
quasiparticle weight is found to decay with a power law $Z \propto
N^{-\nu}$, where the exponent is $\nu = 0.125 \pm 0.001$. This
result indicates that there is a standard orthogonality catastrophe
in the stationary limit.\cite{Anderson-3} For $t_K \rightarrow
\infty$, the dominant term in Eq.~(\ref{eq-1D-H-4}) is $-t_K
\bar{R}$. Since the operator $\bar{R}$ is both unitary and
Hermitian, its eigenvalues are $\pm 1$, and the low-energy subspace
for $t_K \rightarrow \infty$ is the one with $\bar{R} = +1$.
Projecting the remaining terms of $\bar{H}_c (K)$ onto this subspace
with the appropriate operator $(1 + \bar{R}) / 2$, we obtain the
quadratic low-energy Hamiltonian
\begin{eqnarray}
\bar{H}_c (K) &\rightarrow& H_c - \frac{i}{4N} \sum_{k, k'} \Big[
\epsilon_{k'} \left( \bar{\gamma}_{k, \eta, A} + \bar{\gamma}_{k,
\mu, A} \right)
\label{eq-1D-H-5} \\
&& \times \left( \bar{\gamma}_{k', \eta, B} + \bar{\gamma}_{k', \mu,
B} \right) -\epsilon_{k'} \left( \bar{\gamma}_{k, \eta, B} -
\bar{\gamma}_{k, \mu, B} \right)
\nonumber \\
&& \times \left( \bar{\gamma}_{k', \eta, A} - \bar{\gamma}_{k', \mu,
A} \right) \Big] - t_K. \nonumber
\end{eqnarray}
The quasiparticle weight is then found to converge to a large finite
value $Z = | \langle \bar{\omega}_K^{\phantom{\dag}} | \omega
\rangle |^2 \approx 0.97$. This result corroborates our intuition
that the doped ground state is close to the undoped ground state for
$t_K \rightarrow \infty$. In fact, we believe that the ground-state
overlap is only different from $1$ because the modified problem is
equivalent to the original one in terms of the lowest-energy
fermions only. Furthermore, since there is a low-energy
delta-function peak with $Z \sim 1$ in the spectral function for
$t_K \rightarrow \infty$, and the spectral function has the same
low-energy functional form for all $t_K > 0$, we deduce that there
is a delta-function peak with a finite quasiparticle weight $Z > 0$
for all hopping amplitudes $t > 0$ and all hole momenta $K \ll 1$.
The quasiparticle weight only vanishes in the stationary limit when
the low-energy functional form of the spectral function becomes a
power law.

\section{Discussion} \label{sec-disc}

In the previous two sections, we investigated two complementary
directions to determine whether a mobile hole in the gapless phase
of the Kitaev honeycomb model propagates as a coherent
quasiparticle. In Sec.~\ref{sec-var}, we described a variational
approach for the spatially isotropic point ($J_{x,y,z} = J_0$),
where the model is fully two dimensional (2D). The results apply for
a generic point of the gapless phase because they are robust against
perturbations in $J_{x,y,z}$, and they are valid for both
ferromagnetic and antiferromagnetic couplings because they are
invariant under the transformation $J_{x,y,z} \rightarrow
-J_{x,y,z}$. However, the variational approach is somewhat
uncontrolled as it is not immediately clear how close our trial
state is to the actual ground state of the model. In
Sec.~\ref{sec-1D}, we considered a spatially anisotropic special
point ($J_{x,y} = J_0$ and $J_z = 0$), where the model becomes
effectively one dimensional. In this 1D limit, we used an
asymptotically exact calculation to make definite statements about
the actual ground state. However, the different dimensionality might
correspond to different physics, and it is not immediately clear if
the results are applicable for a generic point of the gapless phase.

To make a connection between the two complementary directions, it is
useful to check what the variational approach gives in the 1D limit.
First, by repeating the steps of Sec.~\ref{sec-stat} with the
density of states $g (\epsilon) \sim 1$ around a Dirac point in one
dimension, we find that the quasiparticle weight can be bounded from
above as $Z \leq \exp [-T_1] \sim N^{-\nu_1}$ in the stationary
limit, where $\nu_1 > 0$ is an unknown exponent. Next, by repeating
the steps of Sec.~\ref{sec-var}, we find that $Z$ remains finite in
the thermodynamic limit for any finite variational parameter
$\varrho > 0$. Due to the relative simplicity of the 1D limit, it is
also possible to make these results more quantitative by determining
that $Z \sim N^{-1/8}$ for $\varrho = 0$ and that $Z \sim
\varrho^{1/4}$ for $\varrho > 0$. The result for $\varrho = 0$ is in
agreement with the numerical exponent $\nu \approx 1/8$ found in
Sec.~\ref{sec-1D}. Furthermore, the variational optimization can be
performed explicitly, and it can be verified that the best
variational parameter is $\varrho \sim (t / J_0)^{4/7}$ in the limit
of $t \ll J_0$ (modulo logarithmic corrections). The details of this
procedure are reported elsewhere.\cite{Halasz-2}

We conclude that the variational approach gives the same results in
the 1D limit as our asymptotically exact calculation in
Sec.~\ref{sec-1D}. The standard orthogonality catastrophe in the
stationary limit is straightforward to understand as the density of
states is constant around the Fermi energy.\cite{Anderson-3}
However, it might be surprising that there is no orthogonality
catastrophe for a mobile hole as a standard orthogonality
catastrophe is typically found for both stationary and mobile holes
in 1D quantum liquids.\cite{1D} This difference with respect to
previous studies is explained by the particle-hole symmetric nature
of the Kitaev spin liquid. As a result of particle-hole symmetry,
the phase shift is an odd function of energy, and therefore it
vanishes at zero energy (i.e., at the Fermi energy).

In fact, this difference between the standard case and the
particle-hole-symmetric case is immediately manifest in the
variational approach if we compare Eqs.~(\ref{eq-stat-sc-1}) and
(\ref{eq-var-sc-3}). In the standard case, a generic perturbation
takes the form of a finite potential $V$ appearing in
Eq.~(\ref{eq-stat-sc-1}). The corresponding form in
Eq.~(\ref{eq-var-sc-3}) for the particle-hole-symmetric case can
then be interpreted as a potential whose strength is linearly
proportional to the fermion energy and therefore vanishes for the
lowest-energy fermions. Also, the same feature appears in the
asymptotically exact calculation, where the factor $\epsilon_{k'}$
in the second term of Eq.~(\ref{eq-1D-H-4}) indicates that the
lower-energy fermions are perturbed less by the presence of the
hole. Despite their different formulations, the two complementary
directions seem to capture the same essential physics.

Since the variational approach is in full agreement with the
asymptotically exact calculation for the 1D limit, we expect that
our variational results for the 2D case in Sec.~\ref{sec-var} are
also valid for the actual ground state at $t > 0$. Furthermore, our
intuition suggests that the hole propagates more coherently in the
2D case than in the 1D limit because the fermions have a smaller
density of states around the Fermi energy. This intuition is
corroborated by the respective orthogonality catastrophes found in
the stationary limit. For the 1D limit, we find a standard
orthogonality catastrophe with a power-law decay, while for the 2D
case, we find a weaker orthogonality catastrophe with a logarithmic
decay. Since we know from Sec.~\ref{sec-1D} that a mobile hole
propagates coherently in the 1D limit, we also anticipate coherent
propagation in the 2D case corresponding to a generic point of the
gapless phase.

We finally address the validity of our results for a generic Kitaev
spin liquid, where $H_K$ in Eq.~(\ref{eq-int-H}) may be perturbed by
generic time-reversal-invariant terms. The low-energy physics is
still captured by a single mode of (dressed) Dirac fermions, but any
local disturbance must be represented by a sum of all local fermion
terms allowed by symmetry.\cite{Song} In this case, the disturbance
due to the hole might not couple to a set of non-degenerate levels
(see Secs.~\ref{sec-stat} and \ref{sec-var}), and therefore the
orthogonality catastrophe might not be governed by a single
(vanishing) phase shift at the Fermi energy. Nevertheless, in the 1D
limit, the levels generically split into independent even and odd
sectors with respect to the mirror-reflection symmetry around the
hole site [see Fig.~\ref{fig-3} and Eq.~(\ref{eq-1D-H-2})]. Since
the levels are non-degenerate within each sector, our earlier
arguments apply and indicate that there is no orthogonality
catastrophe in either sector. Furthermore, there is generically no
orthogonality catastrophe in the 2D case.\cite{Hentschel} For a
finite phase shift at the Fermi energy, the integrand in
Eq.~(\ref{eq-var-T-2}) has two fewer powers of energy, but the
integral is nevertheless finite. We therefore expect that our claims
on coherent propagation remain applicable for a generic Kitaev spin
liquid.

\section{Outlook} \label{sec-out}

In this work, we demonstrated that a single hole propagates as a
coherent quasiparticle in the gapless phase of the Kitaev honeycomb
model. In particular, it was found that the quasiparticle weight $Z$
is finite for any small hopping amplitude $t \ll J_0$ but vanishes
in the stationary limit $t \rightarrow 0$. It is then natural to ask
how the quasiparticle weight scales with the hopping amplitude for
$t \ll J_0$. In the 1D limit, this question is addressed
elsewhere,\cite{Halasz-2} and it is found that $Z \sim (t /
J_0)^{1/7}$, modulo logarithmic corrections. Assuming that
Eq.~(\ref{eq-stat-Z-3}) is a tight upper bound for $Z^2$, with
perhaps a renormalized exponent $\kappa$, we expect by analogy that
the leading-order quasiparticle weight is $Z \sim [\log (J_0 /
t)]^{-\hat{\kappa}}$ in the 2D case, where $\hat{\kappa}$ is an
unknown positive exponent. Nevertheless, it would be instructive to
verify this expectation with a rigorous calculation and determine
the exponents $\kappa$ and $\hat{\kappa}$ in doing so.

The coherent propagation of a single hole suggests that the holes
might form a Fermi liquid at finite doping.\cite{Mei} However, it is
far from obvious whether such a Fermi-liquid state would actually be
stable as hole interactions could be relevant in the gapless phase
and turn the Fermi liquid into some more exotic state. It would
therefore be interesting to develop a controlled approach for
describing the interactions between the holes and discussing the
multi-hole ground state at a small but finite hole density. As a
first step towards achieving this goal, it could be useful to
consider the interactions between two holes in the gapless
phase.\cite{AFM-2} Looking at the various interaction channels, one
could then confirm the Fermi-liquid hypothesis\cite{Mei} or even
find unconventional superconductivity.\cite{You}


\begin{acknowledgments}

We thank R.~Moessner for collaboration on closely related earlier
work [Ref.~\onlinecite{Halasz-1}]. We are also grateful to
L.~I.~Glazman, A.~Kamenev, I.~V.~Lerner, K.~Penc, N.~B.~Perkins, and
S.~H.~Simon for enlightening discussions. G.~B.~H.~is supported by a
fellowship from the Gordon and Betty Moore Foundation (Grant
No.~4304). This work was supported in part by the EPSRC under Grants
No.~EP/I032487/1 and No.~EP/N01930X/1.

\end{acknowledgments}


\appendix

\section{Low-energy sector of the Kitaev honeycomb model} \label{app-1}

The standard solution of the Kitaev honeycomb model\cite{Kitaev}
introduces four Majorana fermions at each site $\mathbf{r}$ and
represents the three spin components as
$\sigma_{\mathbf{r}}^{\alpha} = i \hat{b}_{\mathbf{r}}^{\alpha}
\hat{c}_{\mathbf{r}}$ in terms of these four Majorana fermions
$\hat{b}_{\mathbf{r}}^x$, $\hat{b}_{\mathbf{r}}^y$,
$\hat{b}_{\mathbf{r}}^z$, and $\hat{c}_{\mathbf{r}}$. The undoped
spin Hamiltonian in Eq.~(\ref{eq-int-H}) then becomes
\begin{equation}
H_{\sigma} = \sum_{\alpha} \sum_{\mathbf{r} \in A} J_{\alpha} \big(
i \hat{b}_{\mathbf{r}}^{\alpha} \hat{b}_{\mathbf{r} +
\hat{\mathbf{r}}_{\alpha}}^{\alpha} \big) \big( i
\hat{c}_{\mathbf{r}} \hat{c}_{\mathbf{r} +
\hat{\mathbf{r}}_{\alpha}} \big). \label{eq-app-H-1}
\end{equation}
Since there is a conserved quantity $\hat{u}_{\mathbf{r}, \mathbf{r}
+ \hat{\mathbf{r}}_{\alpha}} \equiv i \hat{b}_{\mathbf{r}}^{\alpha}
\hat{b}_{\mathbf{r} + \hat{\mathbf{r}}_{\alpha}}^{\alpha}$ for each
bond and these conserved quantities all commute with one another,
the model splits into independent bond sectors characterized by
$\hat{u}_{\mathbf{r}, \mathbf{r} + \hat{\mathbf{r}}_{\alpha}} = \pm
1$. Within each bond sector, the Hamiltonian in
Eq.~(\ref{eq-app-H-1}) is quadratic and hence exactly solvable.
However, there is some subtlety as the Majorana fermion
representation increases the local Hilbert-space dimension from $2$
to $4$ at each site $\mathbf{r}$. All physical states then satisfy a
corresponding local constraint $D_{\mathbf{r}} \equiv
\hat{b}_{\mathbf{r}}^x \hat{b}_{\mathbf{r}}^y \hat{b}_{\mathbf{r}}^z
\hat{c}_{\mathbf{r}} = 1$ in the Majorana fermion representation,
which acts as a local gauge transformation $\hat{u}_{\mathbf{r},
\mathbf{r} + \hat{\mathbf{r}}_{\alpha}} \rightarrow
-\hat{u}_{\mathbf{r}, \mathbf{r} + \hat{\mathbf{r}}_{\alpha}}$ at
the three bonds around the site $\mathbf{r}$. Importantly, unlike
the bond operators $\hat{u}_{\mathbf{r}, \mathbf{r} +
\hat{\mathbf{r}}_{\alpha}}$ themselves, their product $W_C = \prod_C
\hat{u}_{\mathbf{r}, \mathbf{r} + \hat{\mathbf{r}}_{\alpha}}$ is
gauge invariant around any closed loop $C$ of the lattice, and the
flux operators $W_C$ are then identified as corresponding to gapped
flux excitations. Indeed, it can be shown\cite{Kitaev, Lieb} that
the ground state of the model is in the zero-flux sector
characterized by $W_C = +1$ for all $C$ and that any flux excitation
$W_C = -1$ costs a finite energy $\Delta \sim J_{x,y,z}$.

For a small enough hopping amplitude $t \ll \Delta$, we can restrict
our attention to the low-energy sector with no flux excitations and
represent this zero-flux sector with the trivial bond sector
characterized by $\hat{u}_{\mathbf{r}, \mathbf{r} +
\hat{\mathbf{r}}_{\alpha}} = +1$ for all bonds. The undoped spin
Hamiltonian in Eq.~(\ref{eq-app-H-1}) then immediately reduces to
the corresponding fermion Hamiltonian in Eq.~(\ref{eq-gen-H-3}).
Furthermore, in the Majorana fermion representation, the diagonal
blocks of the doped spin Hamiltonian in Eq.~(\ref{eq-gen-H-2}) are
\begin{equation}
\tilde{H}_{\sigma} (\mathbf{r}, \mathbf{r}) = \left[ H_{\sigma} -
\sum_{\alpha} J_{\alpha} \big( i \hat{b}_{\mathbf{r}}^{\alpha}
\hat{b}_{\mathbf{r} \pm \hat{\mathbf{r}}_{\alpha}}^{\alpha} \big)
\big( i \hat{c}_{\mathbf{r}} \hat{c}_{\mathbf{r} \pm
\hat{\mathbf{r}}_{\alpha}} \big) \right], \label{eq-app-H-2}
\end{equation}
while its non-vanishing off-diagonal blocks are
\begin{equation}
\tilde{H}_{\sigma} (\mathbf{r}, \mathbf{r} \pm
\hat{\mathbf{r}}_{\alpha}) = -\frac{t}{2} \left[ 1 - \sum_{\alpha'}
\big( i \hat{b}_{\mathbf{r}}^{\alpha'} \hat{b}_{\mathbf{r} \pm
\hat{\mathbf{r}}_{\alpha}}^{\alpha'} \big) \big( i
\hat{c}_{\mathbf{r}} \hat{c}_{\mathbf{r} \pm
\hat{\mathbf{r}}_{\alpha}} \big) \right], \label{eq-app-H-3}
\end{equation}
where the upper (lower) sign in front of $\hat{\mathbf{r}}_{\alpha}$
corresponds to $\mathbf{r} \in A$ ($\mathbf{r} \in B$). Since the
terms with $\alpha' \neq \alpha$ in Eq.~(\ref{eq-app-H-3}) create
flux excitations, they have vanishing matrix elements within the
zero-flux sector.\cite{Halasz-1} Neglecting these terms, and using
$i \hat{b}_{\mathbf{r}}^{\alpha} \hat{b}_{\mathbf{r} -
\hat{\mathbf{r}}_{\alpha}}^{\alpha} = -i \hat{b}_{\mathbf{r} -
\hat{\mathbf{r}}_{\alpha}}^{\alpha} \hat{b}_{\mathbf{r}}^{\alpha}$
in the case of $\mathbf{r} \in B$, the blocks of the fermion
Hamiltonian in Eq.~(\ref{eq-gen-H-4}) are then recovered.

\section{Fermion-only representation of the spectral function} \label{app-2}

In terms of the position-space electron operators $a_{\mathbf{r},
\sigma}^{\dag}$, the single-hole spectral function in
Eq.~(\ref{eq-gen-A-1}) is
\begin{eqnarray}
\mathcal{A} (\varepsilon, \mathbf{K}) &=& \frac{1}{2N}
\sum_{\lambda} \sum_{\sigma} \sum_{\mathbf{r}, \mathbf{r}'}
\big{\langle} \Omega \big| a_{\mathbf{r}, \sigma}^{\dag} \big|
\tilde{\Phi}_{\lambda} \big{\rangle} \big{\langle}
\tilde{\Phi}_{\lambda} \big| a_{\mathbf{r}',
\sigma}^{\phantom{\dag}} \big| \Omega \big{\rangle}
\nonumber \\
&& \times \, \delta \big[ \varepsilon - \tilde{E}_{\lambda} \big] \,
e^{-i \mathbf{K} \cdot (\mathbf{r} - \mathbf{r}')}. \label{eq-app-A}
\end{eqnarray}
The ground state of the undoped model reads $| \Omega \rangle =
\mathcal{D} | \omega \rangle$ in the Majorana fermion
representation, where $| \omega \rangle$ is the fermion vacuum
state, and $\mathcal{D} \propto \prod_{\mathbf{r}} (1 +
D_{\mathbf{r}})$ is a projection onto the physical subspace with
$D_{\mathbf{r}} = 1$ for all $\mathbf{r}$. Using the hole-spin
picture, the single-hole states $a_{\mathbf{r}, \sigma} | \Omega
\rangle$ are then
\begin{eqnarray}
a_{\mathbf{r}, \uparrow} | \Omega \rangle &=& | \mathbf{r} \rangle
\otimes \left[ \frac{1}{2} \left( 1 + \sigma_{\mathbf{r}}^z \right)
| \Omega \rangle \right]
\nonumber \\
&=& | \mathbf{r} \rangle \otimes \left[ \frac{\mathcal{D}} {2}
\left( 1 + i b_{\mathbf{0}}^z c_{\mathbf{0}} \right) | \omega
\rangle \right],
\label{eq-app-omega} \\
a_{\mathbf{r}, \downarrow} | \Omega \rangle &=& | \mathbf{r} \rangle
\otimes \left[ \frac{1}{2} \, \sigma_{\mathbf{r}}^x \left( 1 -
\sigma_{\mathbf{r}}^z \right) | \Omega \rangle \right]
\nonumber \\
&=& | \mathbf{r} \rangle \otimes \left[ \frac{\mathcal{D}} {2}
\left( i b_{\mathbf{0}}^x c_{\mathbf{0}} \right) \left( 1 - i
b_{\mathbf{0}}^z c_{\mathbf{0}} \right) | \omega \rangle \right].
\nonumber
\end{eqnarray}
After projecting onto the subspaces with $\sigma_{\mathbf{r}}^z =
\pm 1$ in the two cases, respectively, the spin rotation
$\sigma_{\mathbf{r}}^x$ in the second case ensures that the hole
spin is in the $\sigma_{\mathbf{r}}^z = +1$ state. Note that the
Majorana fermions in Eq.~(\ref{eq-app-omega}) are relabeled by their
relative positions with respect to the hole site $\mathbf{r}$ [see
Eq.~(\ref{eq-gen-c})].

Due to the overall translation symmetry, the eigenstates $|
\tilde{\Phi}_{\lambda} \rangle \equiv | \tilde{\Phi}_{\mathbf{K},
\lambda_{\mathbf{K}}} \rangle$ of the doped model are generically
labeled by the hole momentum $\mathbf{K}$ and an additional label
$\lambda_{\mathbf{K}}$. Also, eigenstates with hole momentum
$\mathbf{K}' \neq \mathbf{K}$ do not contribute to the spectral
function $\mathcal{A} (\varepsilon, \mathbf{K})$. Using the
hole-spin picture, and projecting the hole spin into the
$\sigma_{\mathbf{r}}^z = +1$ state in the Majorana fermion
representation, the contributing eigenstates with hole momentum
$\mathbf{K}$ take the forms [see Eq.~(\ref{eq-gen-theta})]
\begin{eqnarray}
\big| \tilde{\Phi}_{\mathbf{K}, \lambda_{\mathbf{K}}}^{\, p = 0}
\big{\rangle} &=& \frac{1} {\sqrt{N}} \left[ \sum_{\mathbf{r} \in A}
e^{i \mathbf{K} \cdot \mathbf{r}} | \mathbf{r} \rangle +
\sum_{\mathbf{r} \in B} e^{i \mathbf{K} \cdot \mathbf{r} + i
\tilde{\vartheta}_{\mathbf{K}, \lambda_{\mathbf{K}}}} | \mathbf{r}
\rangle \right]
\nonumber \\
&& \otimes \left[ \frac{\mathcal{D}} {2} \left( 1 + i
b_{\mathbf{0}}^z c_{\mathbf{0}} \right) \big|
\tilde{\chi}_{\mathbf{K}, \lambda_{\mathbf{K}}}^{\phantom{\dag}}
\big{\rangle} \right],
\label{eq-app-phi} \\
\big| \tilde{\Phi}_{\mathbf{K}, \lambda_{\mathbf{K}}}^{\, p = 1}
\big{\rangle} &=& \frac{1} {\sqrt{N}} \left[ \sum_{\mathbf{r} \in A}
e^{i \mathbf{K} \cdot \mathbf{r}} | \mathbf{r} \rangle +
\sum_{\mathbf{r} \in B} e^{i \mathbf{K} \cdot \mathbf{r} + i
\tilde{\vartheta}_{\mathbf{K}, \lambda_{\mathbf{K}}}} | \mathbf{r}
\rangle \right]
\nonumber \\
&& \otimes \left[ \frac{\mathcal{D}} {2} \left( 1 + i
b_{\mathbf{0}}^z c_{\mathbf{0}} \right) \left( i b_{\mathbf{0}}^x
c_{\mathbf{0}} \right) \big| \tilde{\chi}_{\mathbf{K},
\lambda_{\mathbf{K}}}^{\phantom{\dag}} \big{\rangle} \right].
\nonumber
\end{eqnarray}
There are two degenerate eigenstates $| \tilde{\Phi}_{\mathbf{K},
\lambda_{\mathbf{K}}}^{\, p = 0} \rangle$ and $|
\tilde{\Phi}_{\mathbf{K}, \lambda_{\mathbf{K}}}^{\, p = 1} \rangle$
for each fermion state $| \tilde{\chi}_{\mathbf{K},
\lambda_{\mathbf{K}}}^{\phantom{\dag}} \rangle$, which respectively
correspond to hole quantum numbers $p = 0$ and $p = 1$ in the
language of Ref.~\onlinecite{Halasz-1}. The remaining two quantum
numbers are $h = 0$ and $q = 0$ for all eigenstates in
Eq.~(\ref{eq-app-phi}), even though eigenstates with $h = 1$ might
have lower energies because flux binding is energetically
favorable.\cite{Halasz-1, Willans} Nevertheless, eigenstates with $h
\neq 0$ or $q \neq 0$ have fractional excitations (i.e., fluxes
and/or fermions) bound to the hole and therefore do not contribute
to the spectral function. Using Eqs.~(\ref{eq-app-omega}) and
(\ref{eq-app-phi}), the matrix elements in Eq.~(\ref{eq-app-A}) are
given by
\begin{eqnarray}
&& \big{\langle} \tilde{\Phi}_{\mathbf{K}, \lambda_{\mathbf{K}}}^{\,
p = 0} \big| a_{\mathbf{r} \in A, \uparrow} \big| \Omega
\big{\rangle} = \big{\langle} \tilde{\Phi}_{\mathbf{K},
\lambda_{\mathbf{K}}}^{\, p = 1} \big| a_{\mathbf{r} \in A,
\downarrow} \big| \Omega \big{\rangle}
\nonumber \\
&& \qquad \qquad \qquad \qquad \, = \frac{1} {2 \sqrt{N}} \,
\big{\langle} \tilde{\chi}_{\mathbf{K},
\lambda_{\mathbf{K}}}^{\phantom{\dag}} \big| \omega \big{\rangle} \,
e^{-i \mathbf{K} \cdot \mathbf{r}},
\nonumber \\
&& \big{\langle} \tilde{\Phi}_{\mathbf{K}, \lambda_{\mathbf{K}}}^{\,
p = 0} \big| a_{\mathbf{r} \in B, \uparrow} \big| \Omega
\big{\rangle} = \big{\langle} \tilde{\Phi}_{\mathbf{K},
\lambda_{\mathbf{K}}}^{\, p = 1} \big| a_{\mathbf{r} \in B,
\downarrow} \big| \Omega \big{\rangle}
\label{eq-app-phi-omega} \\
&& \qquad \qquad \qquad \qquad \, = \frac{1} {2 \sqrt{N}} \,
\big{\langle} \tilde{\chi}_{\mathbf{K},
\lambda_{\mathbf{K}}}^{\phantom{\dag}} \big| \omega \big{\rangle} \,
e^{-i \mathbf{K} \cdot \mathbf{r} - i \tilde{\vartheta}_{\mathbf{K},
\lambda_{\mathbf{K}}}},
\nonumber \\
&& \big{\langle} \tilde{\Phi}_{\mathbf{K}, \lambda_{\mathbf{K}}}^{\,
p = 0} \big| a_{\mathbf{r}, \downarrow} \big| \Omega \big{\rangle} =
\big{\langle} \tilde{\Phi}_{\mathbf{K}, \lambda_{\mathbf{K}}}^{\, p
= 1} \big| a_{\mathbf{r}, \uparrow} \big| \Omega \big{\rangle} = 0.
\nonumber
\end{eqnarray}
Substituting Eq.~(\ref{eq-app-phi-omega}) into Eq.~(\ref{eq-app-A}),
and summing over $p$, the single-hole spectral function in
Eq.~(\ref{eq-gen-A-2}) is then recovered.



\end{document}